\documentclass[preprint,nofootinbib,3p,showpacs]{elsarticle}
\usepackage{graphicx}
\usepackage{amsmath}
\usepackage{amssymb}
\usepackage{overpic,subfigure}
\usepackage{multirow}
\usepackage[symbol]{footmisc}
\usepackage{dcolumn}
\usepackage{bm}
\usepackage{rotating}
\usepackage{hyperref}
\hypersetup{colorlinks=true}
\usepackage{indentfirst}
\usepackage{lineno}
\usepackage{epstopdf}
\usepackage{units}
\usepackage{amsthm}
\usepackage{url}
\usepackage{amsfonts}
\usepackage{textcomp}
\usepackage{multicol}
\usepackage{verbatim}
\usepackage{rotating}
\usepackage{float}
\usepackage{color}
\usepackage{pstricks}
\usepackage{pst-node}
\usepackage{times}
\usepackage{ulem}
\usepackage{booktabs}
\usepackage{multicol}

\def \ee   {e^+e^-}

\def \gev  {\mbox{GeV}}

\newcommand{\jpsi}{J/\psi}

\newcommand{\bfg}{\begin{figure}}
\newcommand{\efg}{\end{figure}}
\newcommand{\bitm}{\begin{itemize}}
\newcommand{\eitm}{\end{itemize}}
\newcommand{\bnum}{\begin{enumerate}}
\newcommand{\enum}{\end{enumerate}}
\newcommand{\btbl}{\begin{table}}
\newcommand{\etbl}{\end{table}}
\newcommand{\btbu}{\begin{tabular}}
\newcommand{\etbu}{\end{tabular}}

\newcommand{\beq}{\begin{equation}}
\newcommand{\edq}{\end{equation}}


\begin{document}
\begin{frontmatter}
\title{\boldmath Search for the lepton flavor violating decay $\jpsi\to e\mu$}

\author{BESIII collaboration\fnref{authors} \corref{email}}
\fntext[authors]{Authors are listed at the end of this paper.}
\cortext[email]{Corresponding email: besiii-publications@ihep.ac.cn}

\date{\today}

\begin{abstract}

We present a search for the lepton flavor violating decay $\jpsi\to e^{\pm}\mu^{\mp}$ using $8.998\times 10^{9} \jpsi$ events collected with the BESIII detector at the BEPCII $e^+e^-$ storage ring. No excess of signal above background is observed;  we therefore set an upper limit on the branching fraction of $\mathcal{B}(\jpsi\to e^{\pm}\mu^{\mp})<4.5\times10^{-9}$ at the $90\%$ confidence level. Improving the previous best result by a factor of more than 30, this measurement places the most stringent limit to date on lepton flavor violation in the heavy quarkonium sector.

\end{abstract}

\end{frontmatter}


\section{Introduction}
\label{sec:introduction}
The discovery of neutrino oscillations and hence non-zero neutrino masses shows that lepton flavor violating (LFV) processes happen in the neutrino sector. In the Standard Model (SM), direct transitions inducing charged lepton flavor violation (CLFV) are absent, and CLFV processes induced by neutrino oscillations are suppressed to levels that are unmeasurable in any current experiment. No CLFV process has been observed so far. The observation of any CLFV process would be a clear signal of new physics beyond the SM~\cite{Bern:2013, Cei:2014}.

In the past decades, the search for CLFV has attracted great interest since it has the potential to probe new physics at energy scales much higher than what is going to be accessible by colliders in the foreseeable future~\cite{Calibbi:2017uvl,pdg:2022}.

Theoretically, CLFV is described by various new physics models, such as the two Higgs doublet model with extra Yukawa couplings~\cite{Hou:2021zqq}, a model that constitutes a simple example of tree-level off-diagonal Majorana couplings not suppressed by neutrino masses~\cite{Escribano:2021uhf}. Another class of models relies on Supersymmetry (SUSY), including SUSY-based Grand Unified Theories (GUT)~\cite{Dimo:1981}, SUSY with vector-like leptons~\cite{Kitano:2000}, SUSY with a right-handed neutrino~\cite{Borzu:1986}, models with a $Z^{'}$~\cite{Berna:1993}, etc. Meanwhile, recent evidence for an anomalous magnetic moment of the muon by the Fermilab $g-2$ experiment \cite{Muong-2:2021ojo} may result from flavor changing neutral couplings to exotic neutral bosons in the two-Higgs doublet model \cite{Hou:2021qmf}. Within this approach, the reaction rate of $\mu\to e\gamma$ is predicted to be within the sensitivity of the MEG~II experiment. The $\mu e \gamma$ dipole can be probed further by $\mu\to 3e$~\cite{Arndt:2021mu3etdr} and $\mu N\to eN$~\cite{Blondel:2013ia,Kuno:1999jp}. The measurements of the branching fractions to electrons and muons involving b to s transitions could be evidence of a new fundamental force with a preferential coupling to muons~\cite{CAMALICH20221}. 
For the CLFV decays of $\jpsi$, models involving GUTs~\cite{Huang:2018guts}, SUSY~\cite{Chang:2000sysu}, and TC2~\cite{Cvetic:1999tc} with various new bosons (scalars, vectors) predict decay rates up to a detectable level of around $10^{-8} \sim 10^{-16}$, possibly providing evidence of new physics~\cite{Nussinov:2000nm,Gutsche:2011bi,Bordes:2000gd,Sun:2012hb,Hazard:2016fnc,Dong:2017ipa}.

Experimentally, massive data sets have been analyzed to study possible CLFV decays of leptons ($\mu, \tau$), pseudoscalar mesons ($K, \pi, B$), and vector mesons ($\phi, \jpsi, \Upsilon$). An upper limit (UL) on the branching fraction (BF) of the $\mu^{+}\to e^{+}\gamma$ decay has been found at $\mathcal{B}(\mu^{+}\to e^{+}\gamma)<4.2\times10^{-13}$ at the $90\%$ confidence level (C.L.) by the MEG Collaboration~\cite{meg:2016}; the limit on the BF of the similar $\tau$ decay has been set by the BABAR Collaboration to $\mathcal{B}(\tau^{+}\to e^{+}\gamma)<3.3\times10^{-8}$~\cite{babar:2010}. A strong limit of $\mathcal{B}(\mu \to 3e)<1.0\times10^{-12}$ has been set by SINDRUM~\cite{sind1988}. The KTeV Collaboration, BNL E871 Collaboration and LHCb collaboration have reported the results $\mathcal{B}(\pi^{0}\to e^{\pm}\mu^{\mp})<3.6\times10^{-10}$~\cite{KTeV:2008}, $\mathcal{B}(K_{L}^{0}\to e^{\pm}\mu^{\mp})<4.7\times10^{-12}$~\cite{BNL:1998} and $\mathcal{B}(B^{0}\to e^{\pm}\mu^{\mp})<1.0\times10^{-9}$~\cite{lhcb:2018} for neutral pion, kaon and $B^{0}$ meson decays, respectively. The LHC ATLAS Collaboration has recently put a strong bound on the $Z\to e^{\pm}\mu^{\mp}$ decay mode. Currently, the best (95$\%$ C.L.) limits are: $\mathcal{B}(Z\to   e^{\pm}\mu^{\mp})<7.5\times10^{-7}$, $\mathcal{B}(Z\to   e^{\pm}\tau^{\mp})<9.8\times10^{-6}$, and $\mathcal{B}(Z\to   {\mu}^{\pm}\tau^{\mp})<1.2\times10^{-5}$~\cite{ATLAS:2014vur,OPAL:1995grn,DELPHI:1996iox}. The current upper bounds for CLFV in Higgs decays are: $\mathcal{B}(H^{0}\to e^{\pm}\mu^{\mp})<6.1\times10^{-5}$, $\mathcal{B}(H^{0}\to e^{\pm}\tau^{\mp})<4.7\times10^{-3}$, and $\mathcal{B}(H^{0}\to   {\mu}^{\pm}\tau^{\mp})<2.5\times10^{-3}$~\cite{ATLAS:2019pmk,CMS:2017con,ATLAS:2019old}.

For quarkonium CLFV decays, using $8.5~\rm{pb}^{-1}$ of $\ee$ annihilation data collected at center-of-mass energies $\sqrt{s}=984-1060~\rm{MeV}$, the most stringent limit of $\mathcal{B}(\phi\to e^{\pm}\mu^{\mp})<2\times10^{-6}$~\cite{SND:2010} has been obtained by the SND Collaboration for the $\phi(1020)$ meson decay.
In addition, the best limits for bottomonium have been reported by the CLEO Collaboration with $\mathcal{B}(\Upsilon(1S, 2S, 3S)\to \mu^{\mp}\tau^{\pm})<\mathcal{O}(10^{-6})$~\cite{cleo3:2008} and the UL on the BF of $\Upsilon(3S)\to e^{\pm}\mu^{\mp}$ decay has been found to be $\mathcal{B}(\Upsilon(3S)\to   e^{\pm}\mu^{\mp})<3.6\times10^{-7}$ by the BABAR experiment~\cite{BaBar:2021loj}. Recently, CLFV decays of $\Upsilon(1S)$ have been investigated by the Belle experiment with $\mathcal{B}(\Upsilon(1S)\to e^{\pm}\mu^{\mp})<3.6\times10^{-7}$, $\mathcal{B}(\Upsilon(1S)\to \mu^{\pm}\tau^{\mp})<2.6\times10^{-6}$, and $\mathcal{B}(\Upsilon(1S)\to e^{\pm}\tau^{\mp})<2.4\times10^{-6}$~\cite{belle:2021}.
In the charmonium sector, the BES experiment has found that $\mathcal{B}(\jpsi\to \mu^{\pm}\tau^{\mp})<2.0\times10^{-6}$ by analyzing $58\times10^{6}~\jpsi$ events~\cite{bes:2004}. Finally, the best ULs on the BFs of the decay $\jpsi\to e^{\pm}\mu^{\mp}$ and $\jpsi\to e^{\pm}\tau^{\mp}$ have been measured at BESIII at the 90$\%$ C.L., with $\mathcal{B}(\jpsi\to e^{\pm}\mu^{\mp})<1.6\times10^{-7}$~\cite{bes3:2013} and $\mathcal{B}(\jpsi\to e^{\pm}\tau^{\mp})<7.1\times10^{-8}$~\cite{BESIII:2021slj}. 

In this paper, we report a search for the CLFV process $\jpsi\to e^{\pm}\mu^{\mp}$ using $8.998\times 10^{9}$ $\jpsi$ events collected with the BESIII detector~\cite{Ablikim:2009aa}. For simplicity, we use the notation $\jpsi\to e\mu$ to represent $\jpsi\to e^{\pm}\mu^{\mp}$ throughout this paper. In Sect.~\ref{sec:detector}, we describe the BESIII detector and the data samples used in this analysis. We introduce the event selection method in Sect.~\ref{sec:event} and present the study of background contributions in Sect.~\ref{sec:background}, followed by systematic uncertainties in Sect.~\ref{sec:systematic}. Finally, we present our results in Sect.~\ref{sec:result} and provide a summary in Sect.~\ref{sec:summary}. 

\section{BESIII Detector and data samples}
\label{sec:detector}

The BESIII detector records symmetric $e^+e^-$ collisions provided by the BEPCII storage ring~\cite{Yu:IPAC2016-TUYA01}, which operates in the center-of-mass energy range from 2.0 to 4.7~GeV. BESIII has collected large data samples in this energy region~\cite{BESIII:2020nme}. The cylindrical core of the BESIII detector covers 93\% of the full solid angle and consists of a helium-based multilayer drift chamber~(MDC), a plastic scintillator time-of-flight system~(TOF), and a CsI(Tl) electromagnetic calorimeter~(EMC), which are all enclosed in a superconducting solenoidal magnet providing a 1.0~T magnetic field. The solenoid is supported by an octagonal flux-return yoke with resistive plate counter muon identification modules interleaved with steel~(MUC). The charged-particle momentum resolution at $1~{\rm GeV}/c$ is $0.5\%$, and the ${\rm d}E/{\rm d}x$ resolution is $6\%$ for electrons from Bhabha scattering. Photon energy is measured by the EMC with a resolution of $2.5\%$ ($5\%$) at $1$~GeV in the barrel (end cap) region. The time resolution in the TOF barrel region is 68~ps, while that in the end cap region is 110~ps. The end cap TOF system was upgraded in 2015 using multi-gap resistive plate chamber technology, providing a time resolution of 60~ps~\cite{etof1, etof2}.

In this paper, $(8.998 \pm 0.040)\times 10^{9}$ $\jpsi$ events collected in 2009, 2018 and 2019 at $\sqrt{s}=3.097~\rm{GeV}$ ~\cite{bes3:totJpsiNumber, bes3:totJpsiNumber2} are analyzed. Monte Carlo (MC) simulated data samples produced with the GEANT4-based~\cite{geant4} package BOOST~\cite{bes:boost}, including the geometry and material description of the BESIII detector~\cite{geo1,geo2,Huang:2022wuo} and the detector response, are used to determine detection efficiencies and to estimate background contributions. $e^+e^-$ collisions collected at $\sqrt{s}=3.773~\gev$, $3.510~\gev$, and $3.080~\gev$~\cite{Ablikim:2013ntc}, corresponding to the integrated luminosities of $2932.8~\rm{pb}^{-1}$, $457.7~\rm{pb}^{-1}$, and $168.6~\rm{pb}^{-1}$, respectively, are used to study the continuum background. The simulation models the beam energy spread and initial state radiation (ISR) in $e^+e^-$ annihilations are modeled with the generator KKMC~\cite{ref:kkmc1, ref:kkmc2}. The inclusive MC sample includes both the production of the $J/\psi$ resonance and the continuum processes incorporated in KKMC. The beam energy and its spread are set according to measurements of BEPCII, and ISR is implemented in the $J/\psi$ generation. EVTGEN~\cite{ref:evtgen1} is used to model the J/psi decays using branching fractions taken from the Particle Data Group~\cite{pdg:2022}, and the remaining unknown charmonium decays are modeled with LUNDCHARM~\cite{Chen:2000tv, ref:lundcharm2}. Final state radiation (FSR) from charged final state particles is incorporated using PHOTOS~\cite{photos}. 

MC simulated samples produced with the GEANT4-based package~\cite{bes3:boss705}, which includes the geometric description of the BESIII detector and the detector response, are used to determine the detection efficiency and to estimate the backgrounds. About 9 billion $\jpsi$ inclusive MC events are used to study the backgrounds from $\jpsi$ decays. For the signal process, 0.3 million $\jpsi\to e\mu$ exclusive events are simulated with PHOTOS and the VLL decay model in EVTGEN. Several kinds of exclusive MC events, including $\jpsi \to e^{+}e^{-}$, $\mu^{+}\mu^{-}$, $\pi^{+}\pi^{-}$, $K^{+}K^{-}$, $p\bar{p}$, and continuum background events $e^{+}e^{-} \to e^{+}e^{-}(\gamma)$, $e^{+}e^{-} \to \mu^{+}\mu^{-}(\gamma)$ are also generated at this energy point to examine the backgrounds.
To avoid involuntary bias during the analysis procedure, a semi-blind analysis method is used in this work. 
About $12.5\%$ of the full $J/\psi$ data are selected randomly for the semi-blind analysis,
and several simulation samples as well as the selected $12.5\%$ data samples are used to optimize the event selection criteria, study the background, and check the reliability of the MC simulation. The full $J/\psi$ data sample is only analyzed after fixing the whole analysis procedure. In this paper, only results with the full data set are shown.

\section{Event Selection}
\label{sec:event}

The selection criteria introduced in the following are optimized by maximizing the figure-of-merit (FOM) according to the signal MC sample and inclusive MC events, given by
\begin{eqnarray}
{\rm{FOM}}=\frac{\epsilon^{\rm MC}_{\rm{sig}}}{\sum_{N^{\rm FOM}_{\rm obs}=0}^{\infty}\mathcal{P}(N^{\rm FOM}_{\rm{obs}}| N^{\rm FOM}_{\rm{exp}})\cdot UL(N^{\rm FOM}_{\rm{obs}}|N^{\rm FOM}_{\rm{exp}})}~.
\label{eq:fom}
\end{eqnarray}
Here, $\epsilon^{\rm MC}_{\rm sig}$ is the detection efficiency from the signal MC sample, $\mathcal{P}(N^{\rm FOM}_{\rm{obs}}|N^{\rm FOM}_{\rm{exp}})$ is the Poisson probability calculated from the number of observed events $N^{\rm FOM}_{\rm{obs}}$ with the expected number of background events $N^{\rm FOM}_{\rm{exp}}$ obtained from inclusive MC samples, and $UL(N^{\rm FOM}_{\rm{obs}}|N^{\rm FOM}_{\rm{exp}})$ is the UL calculated by $N^{\rm FOM}_{\rm{obs}}$ and $N^{\rm FOM}_{\rm{exp}}$ with the Feldman-Cousins method~\cite{feldman:1998ul} at the $90\%$ C.L.~\cite{bes3:2013}.

Each $\jpsi$ signal candidate is reconstructed with only two back-to-back oppositely charged tracks, which would be further identified as one electron and one muon. Charged tracks detected in the MDC are selected with standard criteria first. The tracks are required to be within a polar angle ($\theta$) range of $|\!\cos{\theta}|<0.93$, where $\theta$ is defined with respect to the $z$-axis. They also have to satisfy $R_{xy}<1.0$~cm and $|V_{z}|<10$~cm, where $R_{xy}$ and $|V_{z}|$ are the distances of closest approach to the interaction point of the track in the plane perpendicular to the beam and along the beam direction, respectively.
To reject cosmic rays, the TOF timing difference measured between the two charged tracks must be less than 1.0~ns. In addition, we require the acollinearity angle $|\Delta\theta|=|180^{\circ}-(\theta_{1}+\theta_{2})|<1.2^{\circ}$, where $\theta_{1}$ and $\theta_{2}$ are the polar angles of the two tracks, and the acoplanarity angle $|\Delta\phi|=|180^{\circ}-|\phi_{1}-\phi_{2}||<1.5^{\circ}$, where $\phi_{1}$ and $\phi_{2}$ are the corresponding azimuthal angles, all evaluated in the center-of-mass system of the $e^+ e^-$ collisions.

\begin{figure*}[tbp]
\centering
{\includegraphics[width=0.45\textwidth]{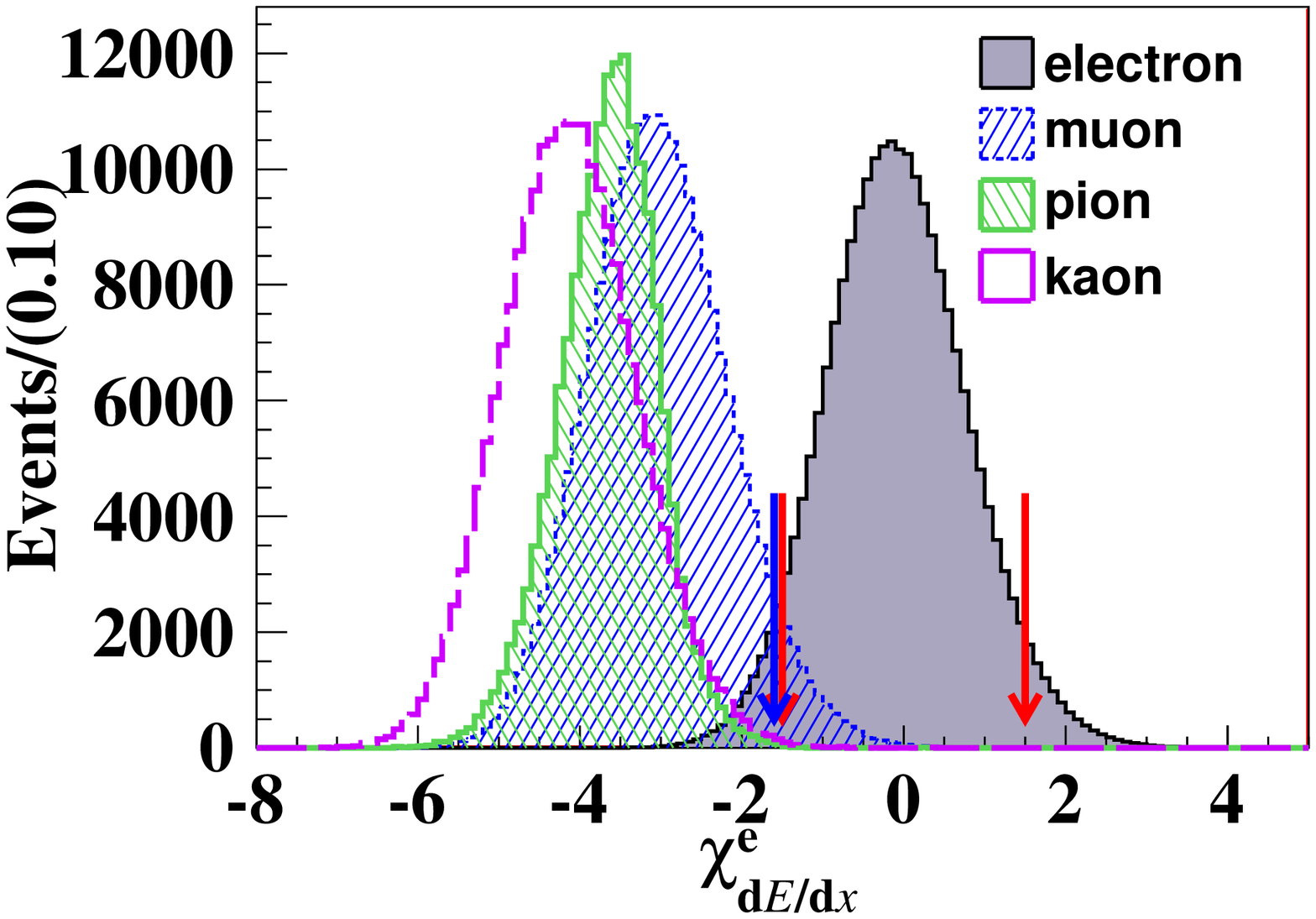}
\hspace{0.03\textwidth}
\includegraphics[width=0.45\textwidth]{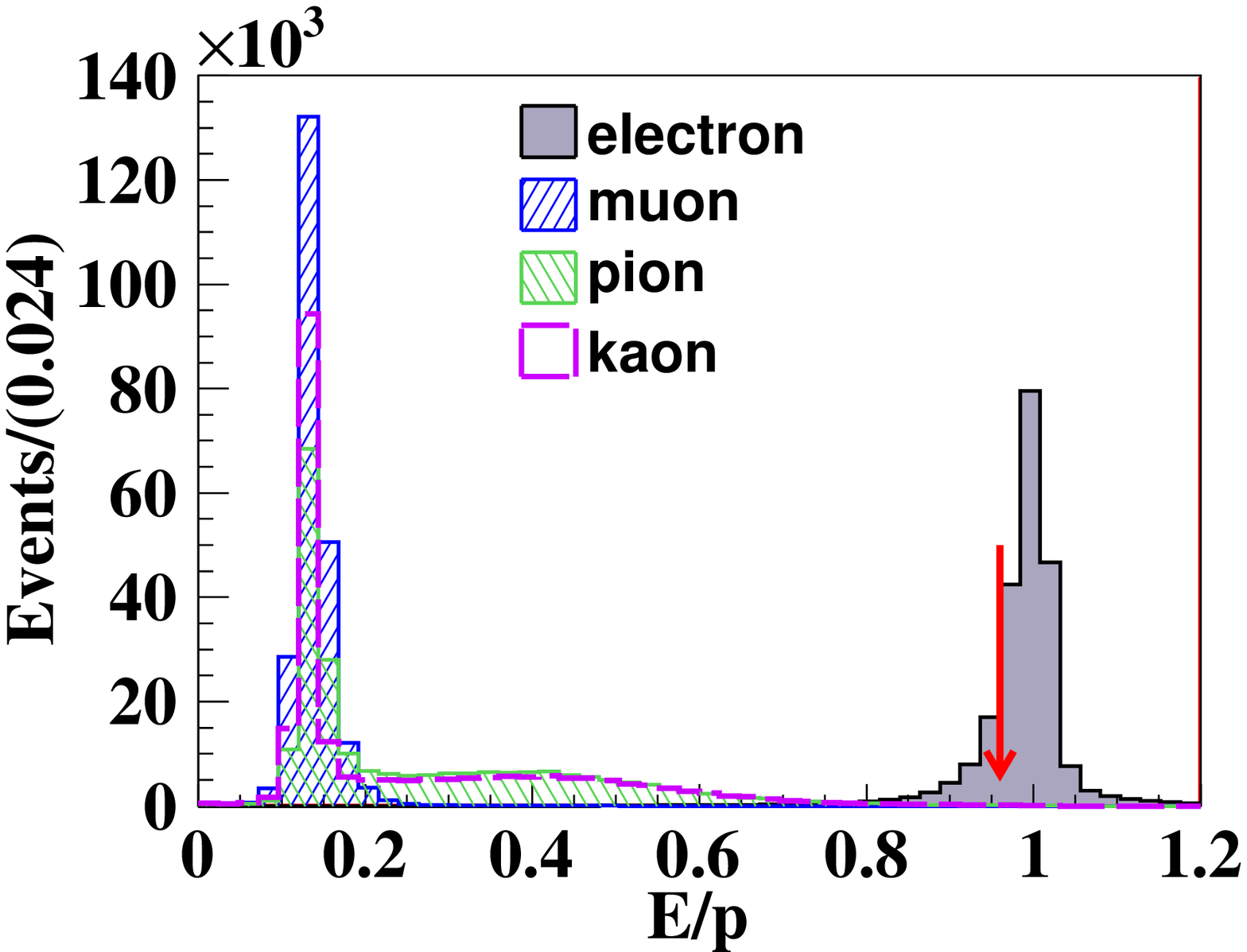}}
\caption{Distributions of $\chi^{e}_{{\rm d}E/{\rm d}x}$ for $e$, $\mu$, $\pi$, and $K$ (left) and $E/p$ for $e$, $\mu$, $\pi$, and $K$ samples (right) from MC simulations of $\jpsi \to e^{+}e^{-}$, $\mu^{+}\mu^{-}$, $\pi^{+}\pi^{-}$, and $K^{+}K^{-}$, respectively. In the left plot the red arrows indicate  $-1.5<\chi^{e}_{{\rm d}E/{ \rm d}x}<1.5$ for electron selection and the blue arrow indicates  $\chi^{e}_{{\rm d}E/{\rm d}x}<-1.6$ for muon selection. In the right plot the red arrow indicates $E/p>0.96$ for electron selection.}
\label{fig:epchi}
\end{figure*}

Electrons and muons are identified based on the information of the MDC, EMC, and MUC sub-detectors.
Electron identification requires no associated hits in the MUC. In addition, the value of $E/p$ is required to be larger than 0.96 for electrons, where $E$ is the deposited energy in the EMC and $p$ is the modulus of the momentum from the MDC. Furthermore, $\chi^{e}_{{\rm d}E/{\rm d}x}$ is required to satisfy $-1.5<\chi^{e}_{{\rm d}E/{\rm d}x}<1.5$, where $\chi^{e}_{{\rm d}E/{\rm d}x}$ is defined as the difference between measured and expected specific energy loss ${\rm d}E/{\rm d}x$ under the electron hypothesis normalized by the ${\rm d}E/{\rm d}x$ resolution.
Fig.~\ref{fig:epchi} shows the distributions of $\chi^{e}_{{\rm d}E/{\rm d}x}$ and $E/p$ for electron, muon, pion, and kaon samples in MC simulated events of $\jpsi \to e^{+}e^{-}$, $\mu^{+}\mu^{-}$, $\pi^{+}\pi^{-}$, and $K^{+}K^{-}$, respectively.

\begin{figure*}[htbp]
\centering
{\includegraphics[width=0.45\textwidth]{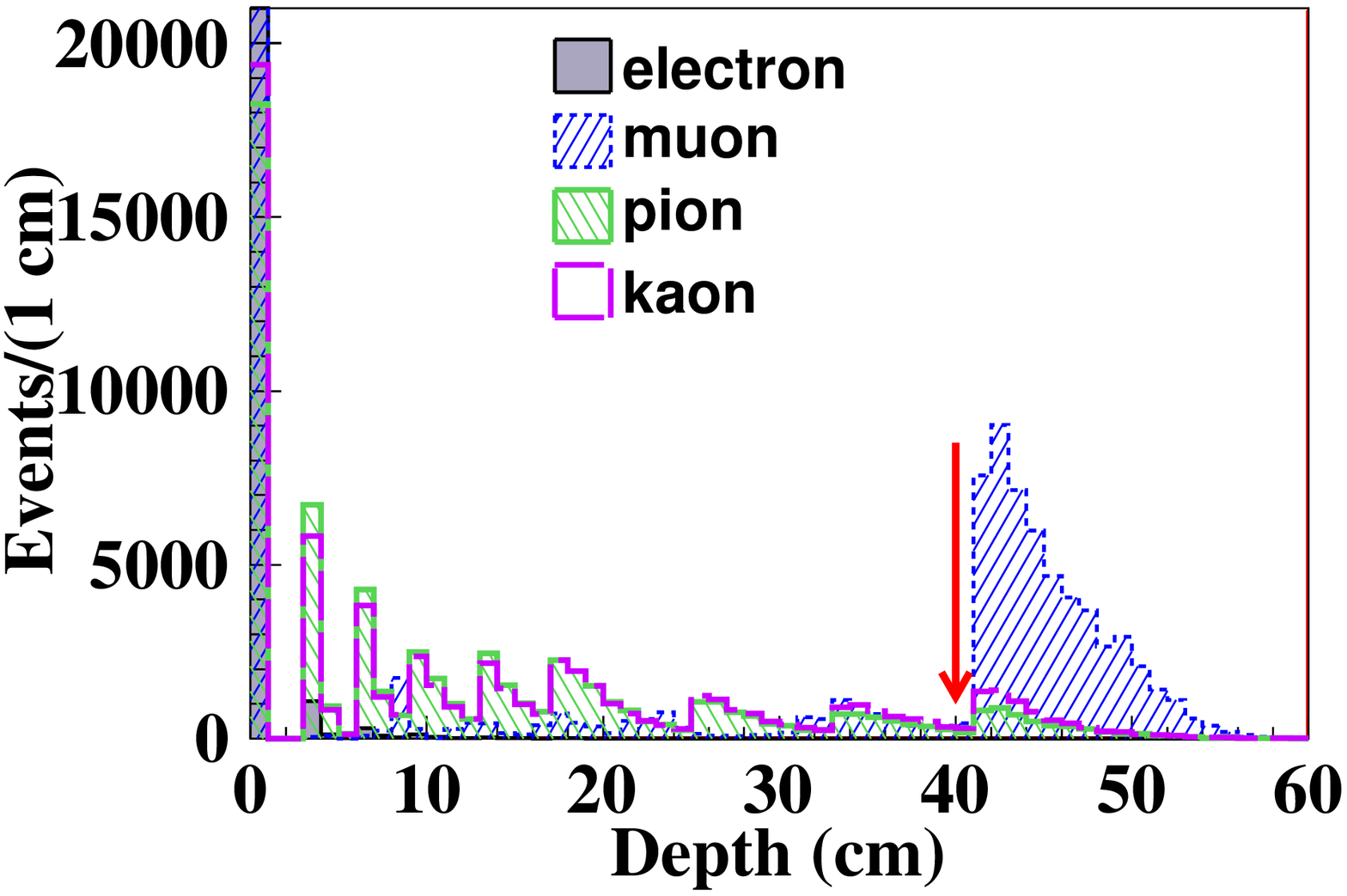}
\includegraphics[width=0.45\textwidth]{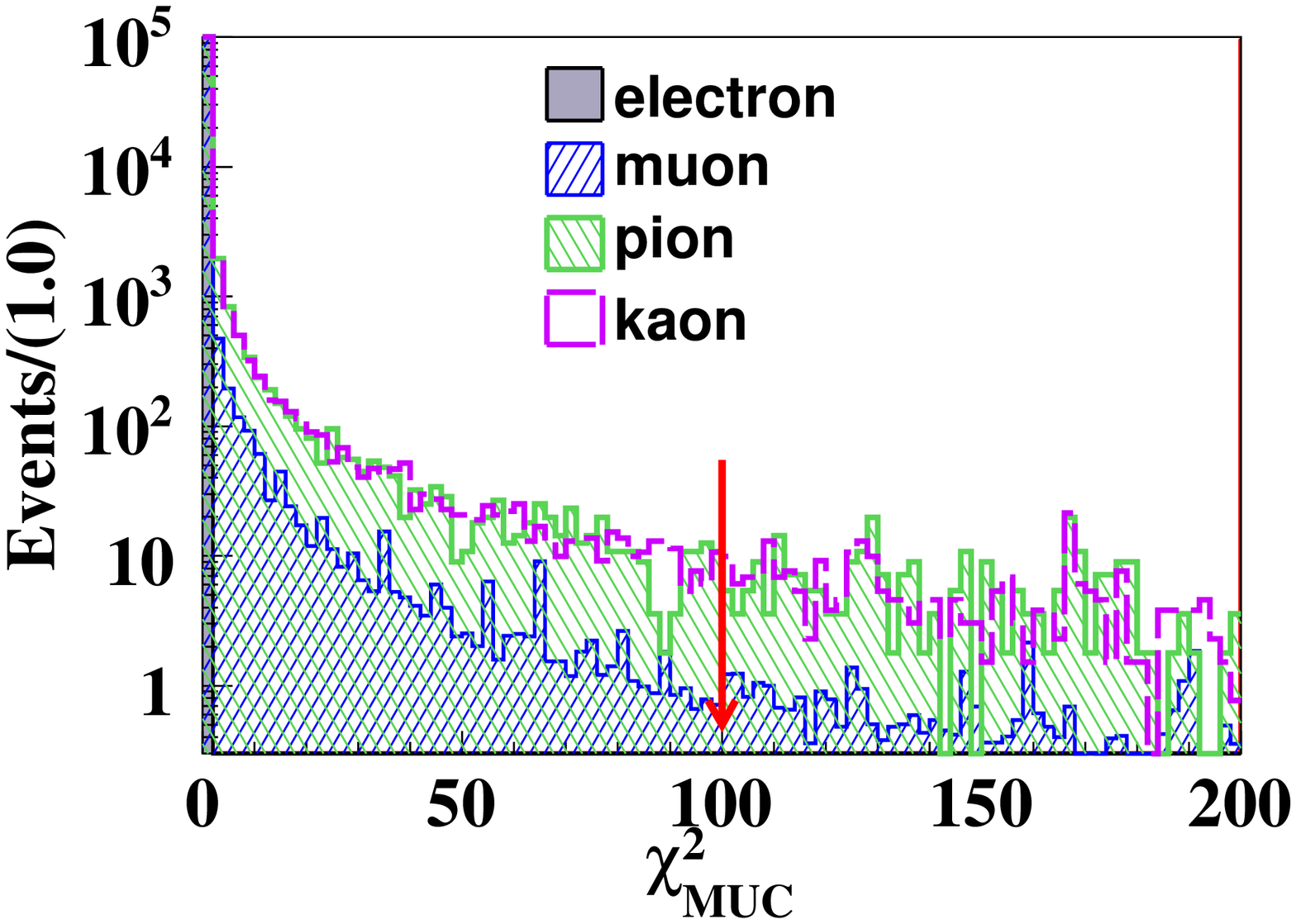}}
\caption{Distributions of the MUC depth (left) and $\chi_{\rm MUC}^{2}$ (right) of different particle samples from MC simulations. The red arrows indicate the muon selection criteria, a MUC depth $>40$~cm in the left plot and $\chi_{\rm MUC}^{2}<100$ in the right plot.}
\label{fig:mucdepth}
\end{figure*}

The muon track is required to have a deposited energy in the EMC within $0.1<E<0.3$~GeV.  To separate muons from other particles, the penetration depth of the track in the MUC is required to be larger than 40~cm. To remove poorly reconstructed MUC tracks and to suppress the background, each muon candidate track must penetrate more than three layers in the MUC and the chi-square value $\chi_{\rm MUC}^2$ of the MUC track is required to be less than 100. The MUC depth and $\chi_{\rm MUC}^{2}$ of different particle samples from MC simulations are shown in Fig.~\ref{fig:mucdepth}, the several jumps on the left is related to different layers of the MUC. Additionally, the $\chi^{e}_{{\rm d}E/{\rm d}x}$ value of a muon candidate must be less than $-1.6$. 

Events with one or more available photon candidates are rejected to suppress the background  from the processes $e^{+}e^{-}\to\gamma e^{+}e^{-}$ and $e^{+}e^{-}\to\gamma \mu^{+}\mu^{-}$.
 Photon candidates are identified using showers in the EMC.  The deposited energy of each shower must be more than 25~MeV in the barrel region ($|\cos \theta|< 0.80$) and more than 50~MeV in the end cap region ($0.86 <|\cos \theta|< 0.92$).  To exclude showers that originate from charged tracks, the angle between the position of each shower in the EMC and the closest extrapolated charged track must be greater than 20 degrees. To suppress electronic noise and showers unrelated to the event, the difference between the EMC time and the event start time is required to be within (0, 700)\,ns. 

\vspace{-0.0cm}
\vspace{-0.0cm}

\begin{figure*}[th]
\centering
{\includegraphics[width=0.45\textwidth]{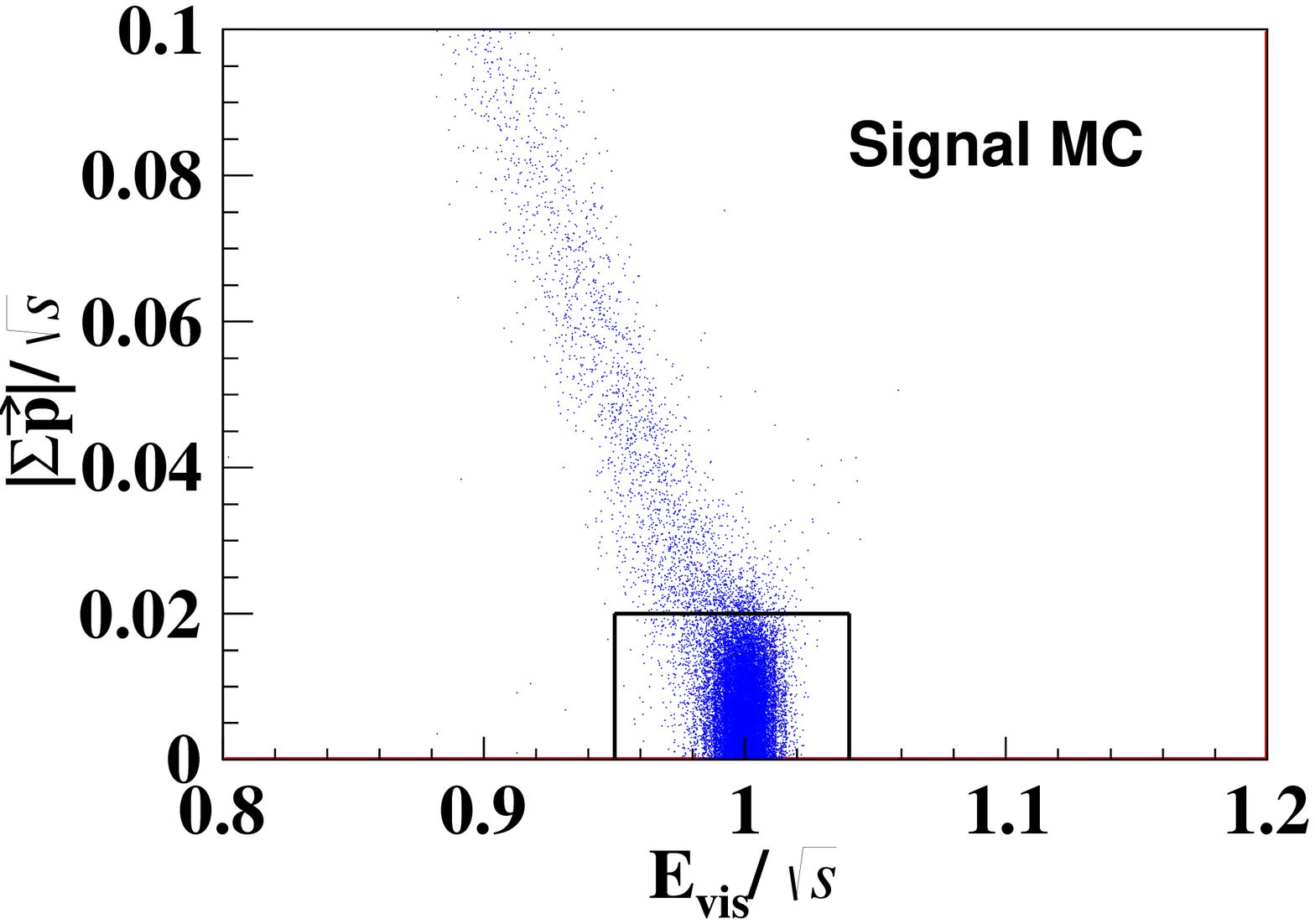}
\hspace{0.1cm}
\includegraphics[width=0.45\textwidth]{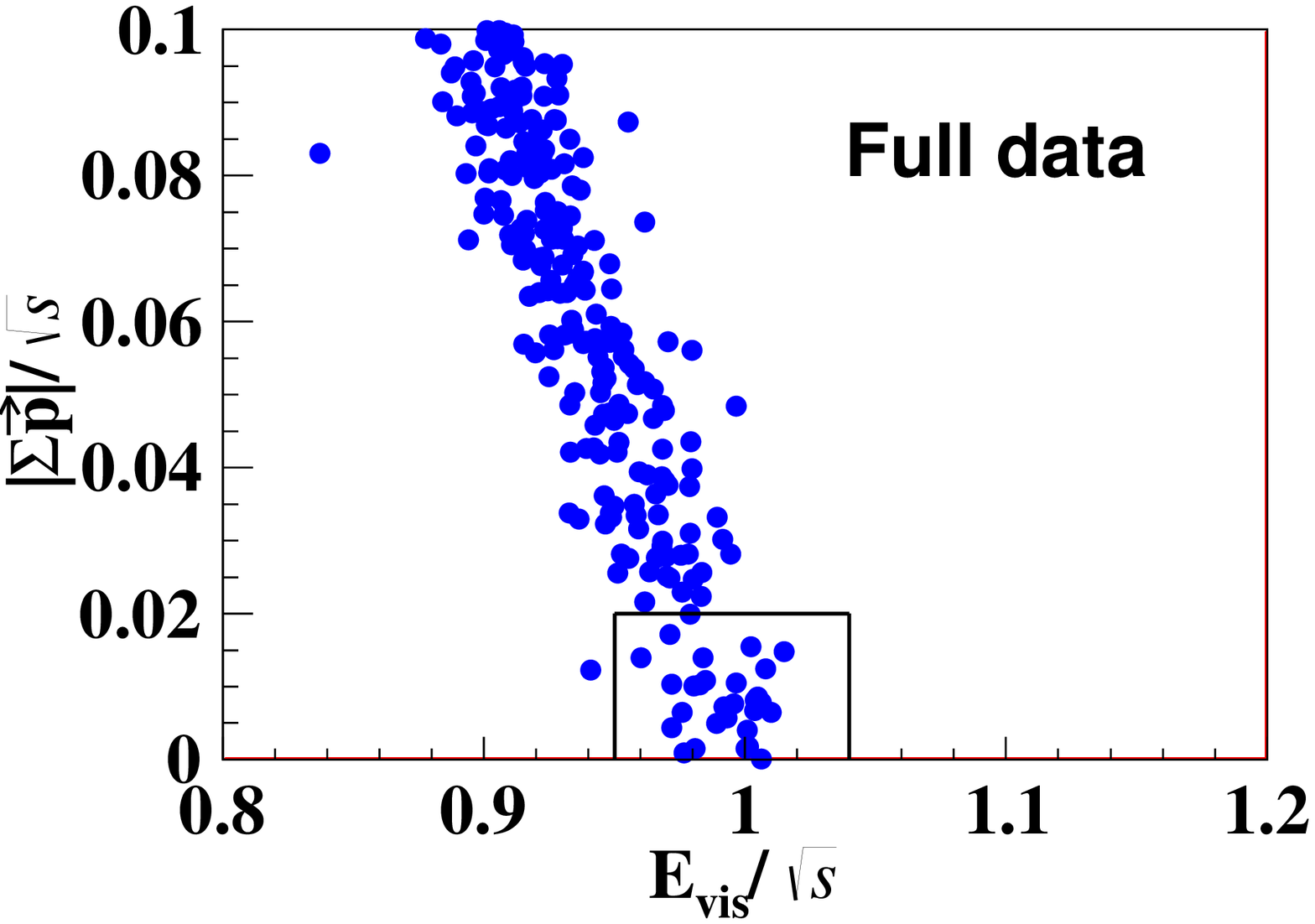}}
\caption{Scatter plots of $|\Sigma\vec{p}|/\sqrt{s}$ versus $E_{\rm vis}/\sqrt{s}$ for the signal MC sample (left) and the $\jpsi$ full data (right). The indicated signal region is defined as $0.95\leqslant E_{\rm vis}/\sqrt{s}\leqslant1.04$ and $|\Sigma\vec{p}|/\sqrt{s}\leqslant0.02$. }
\label{fig:2d_sig}
\end{figure*}

The two variables $|\Sigma\vec{p}|/\sqrt{s}$ and $E_{\rm vis}/\sqrt{s}$ are used to extract the signal yields, where $|\Sigma\vec{p}|$ is the magnitude of the vector sum of the momenta and $E_{\rm vis}$ is the total reconstructed energy of $e$ and $\mu$ in the event. The signal region is defined with $|\Sigma\vec{p}|/\sqrt{s}\leqslant0.02$ and $0.95\leqslant E_{\rm vis}/\sqrt{s}\leqslant1.04$ optimized by maximizing the FOM. As shown in Fig.~\ref{fig:2d_sig}(a), about 85$\%$ of the signal events fall into the signal region.
To avoid possible bias due to the choice of the signal region, we vary the boundary positions slightly by 0.001 and take the most conservative result as the nominal one.

The detection efficiency for the signal process is determined to be $(21.18\pm0.13)\%$. Based on the total $8.998\times 10^{9}$ $\jpsi$ events, $29$ candidate events are observed in the signal region for the $\jpsi\to e\mu$ decays, as shown in Fig.~\ref{fig:2d_sig}(b). 

\vspace{-0.0cm}

\section{Background study}
\label{sec:background}

Two types of background events contaminate the signal region. One results from $\jpsi$ decaying into two charged particle tracks, including $\jpsi \to e^{+}e^{-}$, $\mu^{+}\mu^{-}$, $\pi^{+}\pi^{-}$, $K^{+}K^{-}$, $p\bar{p}$. The dominant background contribution hereby comes from $e-\mu$ misidentification in $\jpsi \to e^{+}e^{-}$ and $\jpsi \to \mu^{+}\mu^{-}$. The other type is continuum background from $\ee$ annihilations into pairs of charged particles, such as $e^{+}e^{-} \to e^{+}e^{-}(\gamma)$, $\mu^{+}\mu^{-}(\gamma)$.

For the $\jpsi$ decay background, we estimate its contributions by analyzing $\jpsi$ inclusive MC events and $\jpsi$ exclusive MC events with about ten times larger statistics compared to the data. The normalized background in the signal region $N^{\rm{norm}}_{\rm{bkg1}}$ is calculated as, 
\begin{equation}
N^{\rm{norm}}_{\rm{bkg1}}=N^{\jpsi-{\rm{MC}}}_{\rm{bkg1}}\cdot f_{1},
~~f_{1}=\frac{N_{\jpsi}^{\rm{data}}}{N_{\jpsi}^{{\rm MC}}}~,
\label{eq:bkg1}
\end{equation}
where $N^{\jpsi-{\rm{MC}}}_{\rm{bkg1}}$ is the number of $\jpsi$ background decays in the $\jpsi$ inclusive and exclusive MC samples, $N_{\jpsi}^{\rm{data}}$ is the total number of $\jpsi$ events in the data, and $N_{\jpsi}^{{\rm{MC}}}$ is the total number of equivalent $\jpsi$ events in the $\jpsi$ inclusive and exclusive MC samples. The yield of background contribution from $\jpsi$ decays in the signal region is estimated to be $N^{\rm{norm}}_{\rm{bkg1}}=24.8\pm1.5$ for the whole $\jpsi$ data. 

The contribution of the continuum background can be estimated using control samples of the $\ee$ collision data at surrounding energy points, such as $\sqrt{s}=3.773$~GeV, $3.510$~GeV, and $3.080$~GeV. By assuming a $1/s$ energy-dependence of the cross sections, the normalized number of continuum backgrounds at the $\jpsi$ peak, $N^{\rm{norm},k}_{\rm{bkg2}}$, can be obtained by
\begin{eqnarray}
N^{\rm{norm},k}_{\rm{bkg2}}=N^{k}_{\rm{cont}}\times f^{k}_2 , ~~f^{k}_2=\frac{\mathcal{L}_{\jpsi}}{\mathcal{L}_{k}}\times\frac{s_{k}}{s_{\jpsi}}~,
\label{eq:normalized1}
\end{eqnarray}
where $N^{k}_{\rm{cont}}$ is the number of background events that have survived in the signal region at the energy with index $k$, while $\mathcal{L}_{k}$ and $\mathcal{L}_{\jpsi}$ are the integrated luminosities at energies $k$ and at the $\jpsi$ peak, respectively. After averaging over the background yields at the three energies, the number of continuum background events is estimated to be $N^{\rm{norm}}_{\rm{bkg2}}=\sum\limits_{k}(N{^{{\rm{norm}},k}_{\rm{bkg2}}}/{\sigma_{k}^2})/{\sum\limits_{k}({1}/{\sigma_{k}^2}})=12.0\pm3.7$, where $\sigma_{k}$ is the error of $N^{{\rm{norm}},k}_{\rm{bkg2}}$.

\section{Systematic uncertainties}
\label{sec:systematic}

Systematic uncertainties arise mainly from efficiency estimation, tracking and PID of electrons and muons, the TOF timing difference, photon veto, and $|\Delta\theta|$ and $|\Delta\phi|$ requirements, while other uncertainties are negligible.
We discuss these major sources of systematic uncertainties in the following.

\begin{itemize}
    \item \emph{Tracking and PID efficiency.} To study the efficiencies of tracking and PID for high-momentum electrons and muons, a subset of $\jpsi$ data and inclusive MC samples is used to select the control samples of $\ee\to\ee$ and $\ee\to\mu^{+}\mu^{-}$. In the control samples, the $e$($\mu$) tracks are selected by applying strict tagging requirements on the oppositely charged $e$($\mu$) tracks to guarantee high purity of the tracks of interest. 
    We observe a relative difference of 0.02$\%$ for the $e$ and $\mu$ tracking efficiencies between data and MC simulations, 3.9$\%$ for the electron PID efficiency and 10$\%$ for the muon PID efficiency, which are taken as systematic uncertainties. Due to the correlation between the two charged tracks in each event, the systematic uncertainties from tracking and PID are considered at the same time and combined. Therefore, the total systematic uncertainty from tracking and PID efficiencies becomes 13$\%$.

    \item \emph{TOF timing.} The $\ee\to\mu^{+}\mu^{-}$ and $\ee\to\ee$ control samples are selected to study the requirement of the TOF timing difference between the two oppositely charged tracks. The relative difference between the efficiencies is $0.19\%$ and $0.48\%$, respectively, and their sum in quadrature, $0.52\%$, is obtained as the corresponding systematic uncertainty.
    
    \item \emph{Photon veto.} The efficiency of the photon veto is studied with control samples of $\ee\to\ee$, $\mu^{+}\mu^{-}$ events in data and MC simulation events, selected as explained for the tracking and PID efficiencies. By comparing the numbers of events with or without photon veto applied in the selected control samples, the selection efficiency can be obtained for electron and muon tracks. The relative differences in data and MC simulations are calculated to be $0.83\%$ and $0.59\%$ for the $\ee$ and $\mu^{+}\mu^{-}$ control samples, respectively. The systematic uncertainty of photon veto for the signal process is then conservatively assigned the maximum of these two, $0.83\%$.
    
    \item \emph{$|\Delta\theta|$ and $|\Delta\phi|$ requirements.} With the same control samples used in the tracking and PID efficiency study, we compare the efficiencies of $|\Delta\theta|$ and $|\Delta\phi|$ requirements for data and MC simulations. The combined systematic uncertainty is calculated to be $2.6\%$ ($2.4\%$) according to the control samples of $\ee$ ($\mu^{+}\mu^{-}$) final states. The systematic uncertainty of $|\Delta\theta|$ and $|\Delta\phi|$ requirements is conservatively assigned the maximum of these two, $2.6\%$.
    
\end{itemize}

All of the above discussed systematic uncertainties are summarized in Table~\ref{tab:syst_err}. They are added in quadrature to the total efficiency-related systematic uncertainty of $14\%$.

\begin{table*}[tpb]
\setlength{\abovecaptionskip}{0.0cm}
\setlength{\belowcaptionskip}{-1.6cm}
\caption{Summary of the efficiency-related relative systematic uncertainties in percent.}
  \begin{center}
  \footnotesize
  \newcommand{\tabincell}[2]{\begin{tabular}{@{}#1@{}}#2\end{tabular}}
  \begin{tabular}{l c c c c c c c}
      \hline\hline
                        Source & Relative uncertainty (\%)\\
			\hline
			Tracking and PID & 13\\
			TOF timing & 0.52 \\
			Photon veto & 0.83\\
            $|\Delta\theta|$ and $|\Delta\phi|$ requirement & 2.6 \\
    			\hline
                         Total & 14\\
      \hline\hline
  \end{tabular}
  \end{center}
  \label{tab:syst_err}
\end{table*}

\section{Results}
\label{sec:result}

We observe 29 candidate signal events in the signal region, while  $36.8\pm4.0$ background events are expected. Hence, no excess is observed and an upper limit on the branching fraction $\mathcal{B}(\jpsi\to e\mu)$ is estimated with the profile likelihood method~\cite{rolke:2005like}. The likelihood function is defined as

\begin{align}
\begin{split}
\mathcal{L}(\mathcal{B}, \epsilon_{\rm{sig}}, N_{\jpsi}, N_{\rm{bkg1}}, N_{\rm{bkg2}}) = &\mathcal{P}(N_{\rm{obs}} |   N_{\jpsi}\cdot\mathcal{B}\cdot\epsilon_{\rm{sig}}+N_{\rm{bkg1}}+N_{\rm{bkg2}})\\
\cdot&\mathcal{G}(\epsilon_{\rm{sig}} |  \epsilon^{\rm{MC}}_{\rm{sig}},  \epsilon^{\rm{MC}}_{\rm{sig}}\cdot\sigma^{\rm{EFF}}_{\rm{sig}})\\
\cdot&\mathcal{P}(N^{\jpsi-{\rm MC}}_{\rm{bkg1}} | N_{\rm{bkg1}}/f_{1})\\
\cdot&\prod \limits_{k}\mathcal{P}(N^{k}_{\rm{cont}} | N_{\rm{bkg2}}/f^{k}_{2})\\
\cdot&\mathcal{G}(N_{\jpsi} |  N_{\jpsi}^{\rm{data}}, \delta N_{\jpsi}^{\rm{data}})~.
\label{eq:likelihood}
\end{split}
\end{align}
Here, $N_{\jpsi}$, $N_{\rm{bkg1}}/f_{1}$ and $N_{\rm{bkg2}}/f^{k}_{2}$ are the expected values of the total $\jpsi$ number in data, of background contributions from $\jpsi$ decays, and from continuum background, respectively. The detection efficiency $\epsilon_{\rm{sig}}$ obeys a Gaussian distribution ($\mathcal{G}$) with mean value of MC-determined $\epsilon^{\rm{MC}}_{\rm{sig}}$ and its absolute uncertainty $\epsilon^{\rm{MC}}_{\rm{sig}}\cdot\sigma^{\rm{EFF}}_{\rm{sig}}$. Here $\sigma^{\rm{EFF}}_{\rm{sig}}$ denotes the sum of the relative statistical and systematic uncertainties on efficiency. 
The numbers of background events, $N^{\jpsi-{\rm MC}}_{\rm{bkg1}}$ in $\jpsi$ inclusive MC samples and $N^{k}_{\rm cont}$ at different continuum energy points, follow a Poisson distribution ($\mathcal{P}$) with expected values of $ N_{\rm{bkg1}}/f_{1}$ and $N_{\rm{bkg2}}/f^{k}_{2}$, respectively, while $N_{\jpsi}$ follows a Gaussian distribution with mean $N^{data}_{\jpsi}$ and standard deviation $\delta N_{\jpsi}^{\rm{data}}$. The values of the parameters in Eq.~\eqref{eq:likelihood} are listed in Table~\ref{tab:value}. 

\begingroup
\renewcommand\arraystretch{1.2}
\begin{table*}[!ht]
\caption{The values of the parameters for the likelihood function in Eq.~\eqref{eq:likelihood}.}
\centering
\footnotesize
\begin{tabular}{lc}
\hline \hline
                    Parameter & Value\\
                    \hline
                    $N_{\rm obs}$ & $29$ \\
                    $N^{\rm data}_{\jpsi}$ & $8.998\times10^{9}$\\
                    $\delta N^{\rm data}_{\jpsi}$ & $0.040\times10^{9}$\\
                    $\epsilon^{\rm MC}_{\rm sig}$ & $21\%$\\
                    $\sigma^{\rm EFF}_{\rm sig}$ & $14\%$\\
                    $N^{\jpsi-{\rm MC}}_{\rm bkg1}$ &
                    $275$\\
                    $N^{3.773}_{\rm cont}$ &$10$ \\
                    $N^{3.510}_{\rm cont}$ &$1$ \\
                    $N^{3.080}_{\rm cont}$ &$0$\\
                    $f_{1}$ & 0.09090\\
                    $f^{3.773}_{2}$ &1.3416\\
                    $f^{3.510}_{2}$&7.4390\\
                    $f^{3.080}_{2}$&15.553\\
                
\hline \hline
\end{tabular}
\label{tab:value}
\end{table*}
\endgroup

\begin{figure*}[htbp]
\centering
\includegraphics[width=0.5\textwidth]{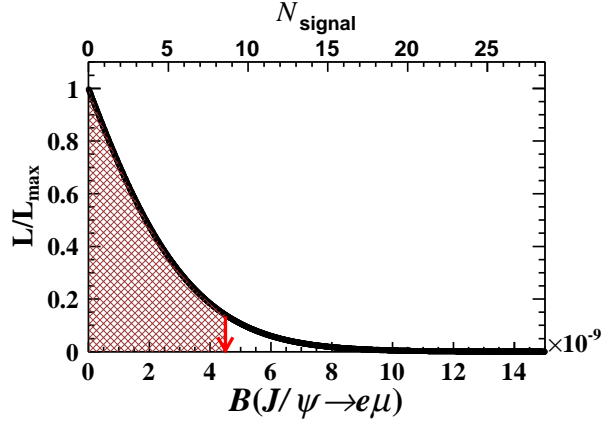}
\caption{Normalized likelihood distribution as a function of the assumed $\mathcal{B}(\jpsi\to e\mu)$. The red arrow points to the position of the UL at the $90\%$ C.L. The upper $x$-axis shows the corresponding values of the signal yields $N_{\rm signal}$. }
\label{fig:likelihood}
\end{figure*}

To estimate the UL of $\mathcal{B}(\jpsi\to e\mu)$, the likelihood function of the scanned branching fraction is obtained. For a sequence of fixed positive values of $\mathcal{B}$, the likelihood values are maximized with respect to all other variables.
The likelihood distribution normalized to the maximum likelihood value $L_{\rm max}$ is shown in Fig.~\ref{fig:likelihood}.
The UL on the BF is found to be $\mathcal{B}(\jpsi\to e\mu)<4.5\times10^{-9}$ at the $90\%$ C.L. by integrating the likelihood curve in the physical region of $\mathcal{B}\geq0$.

\vspace{-0.0cm}
\vspace{-0.0cm}

\section{Summary}
\label{sec:summary}

High statistics $\jpsi$ data provides the opportunity to thoroughly investigate the CLFV $\jpsi\to e^\pm\mu^\mp$ decay to search for new physics.
Based on $8.998\times10^{9} \jpsi$ events collected with the BESIII detector at the BEPCII, a semi-blind analysis is performed and no significant excess is found in the data set with respect to the expected background. The upper limit of the branching fraction is determined to be $\mathcal{B}(\jpsi\to e^\pm\mu^\mp)<4.5\times10^{-9}$ at the $90\%$ C.L., which improves the previous limit by a factor of more than 30. It is currently the most precise result of CLFV search in heavy quarkonium systems and provides constraints on the parameter spaces of new physics models.

\section{Acknowledgments}

This work was supported by the National Key Research and Development Program of China (Grant Nos. 2020YFA
0406400, and 2020YFA0406300), Joint Large-Scale Scientific Facility Funds of the National Natural Science Foundation of China (NSFC) and Chinese Academy of Sciences (CAS) (Grant Nos. U1932101, U1732263, and U1832207), State Key Laboratory of Nuclear Physics and Technology, Peking University (Grant No. NPT2020KFY04), National Natural Science Foundation of China (Grant Nos. 11625523, 11635010, 11675275, 11735014, 11822506, 11835012, 11935015, 11935016, 11935018, 11961141012, 11975021, 12022510, 12035009, 12035013, 12061131003, and 12175321), CAS Center for Excellence in Particle Physics (CCEPP), CAS Large-Scale Scientific Facility Program, CAS Key Research Program of Frontier Sciences (Grant No. QYZDJSSW-SLH040), 100 Talents Program of CAS, Fundamental Research Funds for the Central Universities, Institute of Nuclear and Particle Physics and Shanghai Key Laboratory for Particle Physics and Cosmology, European Research Council (Grant No. 758462), European Union Horizon 2020 Research and Innovation Programme under Contract No. Marie SklodowskaCurie (Grant No. 894790), German Research Foundation DFG (Grant No. 443159800), Collaborative Research Center (Grant Nos. CRC 1044, FOR 2359, and GRK 2149), Istituto Nazionale di Fisica Nucleare, Italy, Ministry of Development of Turkey (Grant No. DPT2006K-120470), National Science and Technology fund, Olle Engkvist Foundation (Grant No. 200-0605), Science and Technology Facilities Council (United Kingdom), The Knut and Alice Wallenberg Foundation (Sweden) (Grant No. 2016.0157), The Royal Society, UK (Grant Nos. DH140054, and DH160214), The Swedish Research Council, U.S. Department of Energy (Grant Nos. DE-FG02-05ER41374,and DE-SC-0012069). The BESIII Collaboration thanks the staff of BEPCII and the IHEP computing center for their strong support.


\bibliographystyle{apsrev4-1}
\bibliography{emu.bib}

\begin{thebibliography}{66}%
\makeatletter
\providecommand \@ifxundefined [1]{%
 \@ifx{#1\undefined}
}%
\providecommand \@ifnum [1]{%
 \ifnum #1\expandafter \@firstoftwo
 \else \expandafter \@secondoftwo
 \fi
}%
\providecommand \@ifx [1]{%
 \ifx #1\expandafter \@firstoftwo
 \else \expandafter \@secondoftwo
 \fi
}%
\providecommand \natexlab [1]{#1}%
\providecommand \enquote  [1]{``#1''}%
\providecommand \bibnamefont  [1]{#1}%
\providecommand \bibfnamefont [1]{#1}%
\providecommand \citenamefont [1]{#1}%
\providecommand \href@noop [0]{\@secondoftwo}%
\providecommand \href [0]{\begingroup \@sanitize@url \@href}%
\providecommand \@href[1]{\@@startlink{#1}\@@href}%
\providecommand \@@href[1]{\endgroup#1\@@endlink}%
\providecommand \@sanitize@url [0]{\catcode `\\12\catcode `\$12\catcode
  `\&12\catcode `\#12\catcode `\^12\catcode `\_12\catcode `\%12\relax}%
\providecommand \@@startlink[1]{}%
\providecommand \@@endlink[0]{}%
\providecommand \url  [0]{\begingroup\@sanitize@url \@url }%
\providecommand \@url [1]{\endgroup\@href {#1}{\urlprefix }}%
\providecommand \urlprefix  [0]{URL }%
\providecommand \Eprint [0]{\href }%
\providecommand \doibase [0]{http://dx.doi.org/}%
\providecommand \selectlanguage [0]{\@gobble}%
\providecommand \bibinfo  [0]{\@secondoftwo}%
\providecommand \bibfield  [0]{\@secondoftwo}%
\providecommand \translation [1]{[#1]}%
\providecommand \BibitemOpen [0]{}%
\providecommand \bibitemStop [0]{}%
\providecommand \bibitemNoStop [0]{.\EOS\space}%
\providecommand \EOS [0]{\spacefactor3000\relax}%
\providecommand \BibitemShut  [1]{\csname bibitem#1\endcsname}%
\let\auto@bib@innerbib\@empty
\bibitem [{\citenamefont {Bernstein}\ and\ \citenamefont
  {Cooper}(2013)}]{Bern:2013}%
  \BibitemOpen
  \bibfield  {author} {\bibinfo {author} {\bibfnamefont {R.~H.}\ \bibnamefont
  {Bernstein}}\ and\ \bibinfo {author} {\bibfnamefont {P.~S.}\ \bibnamefont
  {Cooper}},\ }\href {\doibase 10.1016/j.physrep.2013.07.002} {\bibfield
  {journal} {\bibinfo  {journal} {Phys. Rept.}\ }\textbf {\bibinfo {volume}
  {532}},\ \bibinfo {pages} {27} (\bibinfo {year} {2013})},\ \Eprint
  {http://arxiv.org/abs/1307.5787} {arXiv:1307.5787 [hep-ex]} \BibitemShut
  {NoStop}%
\bibitem [{\citenamefont {Cei}\ and\ \citenamefont {Nicolo}(2014)}]{Cei:2014}%
  \BibitemOpen
  \bibfield  {author} {\bibinfo {author} {\bibfnamefont {F.}~\bibnamefont
  {Cei}}\ and\ \bibinfo {author} {\bibfnamefont {D.}~\bibnamefont {Nicolo}},\
  }\href {\doibase 10.1155/2014/282915} {\bibfield  {journal} {\bibinfo
  {journal} {Adv. High Energy Phys.}\ }\textbf {\bibinfo {volume} {2014}},\
  \bibinfo {pages} {282915} (\bibinfo {year} {2014})}\BibitemShut {NoStop}%
\bibitem [{\citenamefont {Calibbi}\ and\ \citenamefont
  {Signorelli}(2018)}]{Calibbi:2017uvl}%
  \BibitemOpen
  \bibfield  {author} {\bibinfo {author} {\bibfnamefont {L.}~\bibnamefont
  {Calibbi}}\ and\ \bibinfo {author} {\bibfnamefont {G.}~\bibnamefont
  {Signorelli}},\ }\href {\doibase 10.1393/ncr/i2018-10144-0} {\bibfield
  {journal} {\bibinfo  {journal} {Riv. Nuovo Cim.}\ }\textbf {\bibinfo {volume}
  {41}},\ \bibinfo {pages} {71} (\bibinfo {year} {2018})},\ \Eprint
  {http://arxiv.org/abs/1709.00294} {arXiv:1709.00294 [hep-ph]} \BibitemShut
  {NoStop}%
\bibitem [{\citenamefont {Workman}\ and\ \citenamefont
  {Others}(2022)}]{pdg:2022}%
  \BibitemOpen
  \bibfield  {author} {\bibinfo {author} {\bibfnamefont {R.~L.}\ \bibnamefont
  {Workman}}\ and\ \bibinfo {author} {\bibnamefont {Others}} (\bibinfo
  {collaboration} {Particle Data Group}),\ }\href {\doibase
  10.1093/ptep/ptac097} {\bibfield  {journal} {\bibinfo  {journal} {PTEP}\
  }\textbf {\bibinfo {volume} {2022}},\ \bibinfo {pages} {083C01} (\bibinfo
  {year} {2022})}\BibitemShut {NoStop}%
\bibitem [{\citenamefont {Hou}\ \emph {et~al.}(2021)\citenamefont {Hou},
  \citenamefont {Kumar},\ and\ \citenamefont {Teunissen}}]{Hou:2021zqq}%
  \BibitemOpen
  \bibfield  {author} {\bibinfo {author} {\bibfnamefont {W.-S.}\ \bibnamefont
  {Hou}}, \bibinfo {author} {\bibfnamefont {G.}~\bibnamefont {Kumar}}, \ and\
  \bibinfo {author} {\bibfnamefont {S.}~\bibnamefont {Teunissen}},\ }\href@noop
  {} {\  (\bibinfo {year} {2021})},\ \Eprint {http://arxiv.org/abs/2109.08936}
  {arXiv:2109.08936 [hep-ph]} \BibitemShut {NoStop}%
\bibitem [{\citenamefont {Escribano}\ \emph {et~al.}(2021)\citenamefont
  {Escribano}, \citenamefont {Hirsch}, \citenamefont {Nava},\ and\
  \citenamefont {Vicente}}]{Escribano:2021uhf}%
  \BibitemOpen
  \bibfield  {author} {\bibinfo {author} {\bibfnamefont {P.}~\bibnamefont
  {Escribano}}, \bibinfo {author} {\bibfnamefont {M.}~\bibnamefont {Hirsch}},
  \bibinfo {author} {\bibfnamefont {J.}~\bibnamefont {Nava}}, \ and\ \bibinfo
  {author} {\bibfnamefont {A.}~\bibnamefont {Vicente}},\ }\href@noop {} {\
  (\bibinfo {year} {2021})},\ \Eprint {http://arxiv.org/abs/2108.01101}
  {arXiv:2108.01101 [hep-ph]} \BibitemShut {NoStop}%
\bibitem [{\citenamefont {Dimopoulos}\ and\ \citenamefont
  {Georgi}(1981)}]{Dimo:1981}%
  \BibitemOpen
  \bibfield  {author} {\bibinfo {author} {\bibfnamefont {S.}~\bibnamefont
  {Dimopoulos}}\ and\ \bibinfo {author} {\bibfnamefont {H.}~\bibnamefont
  {Georgi}},\ }\href {\doibase 10.1016/0550-3213(81)90522-8} {\bibfield
  {journal} {\bibinfo  {journal} {Nucl. Phys. B}\ }\textbf {\bibinfo {volume}
  {193}},\ \bibinfo {pages} {150} (\bibinfo {year} {1981})}\BibitemShut
  {NoStop}%
\bibitem [{\citenamefont {Kitano}\ and\ \citenamefont
  {Yamamoto}(2000)}]{Kitano:2000}%
  \BibitemOpen
  \bibfield  {author} {\bibinfo {author} {\bibfnamefont {R.}~\bibnamefont
  {Kitano}}\ and\ \bibinfo {author} {\bibfnamefont {K.}~\bibnamefont
  {Yamamoto}},\ }\href {\doibase 10.1103/PhysRevD.62.073007} {\bibfield
  {journal} {\bibinfo  {journal} {Phys. Rev. D}\ }\textbf {\bibinfo {volume}
  {62}},\ \bibinfo {pages} {073007} (\bibinfo {year} {2000})},\ \Eprint
  {http://arxiv.org/abs/hep-ph/0003063} {arXiv:hep-ph/0003063} \BibitemShut
  {NoStop}%
\bibitem [{\citenamefont {Borzumati}\ and\ \citenamefont
  {Masiero}(1986)}]{Borzu:1986}%
  \BibitemOpen
  \bibfield  {author} {\bibinfo {author} {\bibfnamefont {F.}~\bibnamefont
  {Borzumati}}\ and\ \bibinfo {author} {\bibfnamefont {A.}~\bibnamefont
  {Masiero}},\ }\href {\doibase 10.1103/PhysRevLett.57.961} {\bibfield
  {journal} {\bibinfo  {journal} {Phys. Rev. Lett.}\ }\textbf {\bibinfo
  {volume} {57}},\ \bibinfo {pages} {961} (\bibinfo {year} {1986})}\BibitemShut
  {NoStop}%
\bibitem [{\citenamefont {Bernabeu}\ \emph {et~al.}(1993)\citenamefont
  {Bernabeu}, \citenamefont {Nardi},\ and\ \citenamefont
  {Tommasini}}]{Berna:1993}%
  \BibitemOpen
  \bibfield  {author} {\bibinfo {author} {\bibfnamefont {J.}~\bibnamefont
  {Bernabeu}}, \bibinfo {author} {\bibfnamefont {E.}~\bibnamefont {Nardi}}, \
  and\ \bibinfo {author} {\bibfnamefont {D.}~\bibnamefont {Tommasini}},\ }\href
  {\doibase 10.1016/0550-3213(93)90446-V} {\bibfield  {journal} {\bibinfo
  {journal} {Nucl. Phys. B}\ }\textbf {\bibinfo {volume} {409}},\ \bibinfo
  {pages} {69} (\bibinfo {year} {1993})},\ \Eprint
  {http://arxiv.org/abs/hep-ph/9306251} {arXiv:hep-ph/9306251} \BibitemShut
  {NoStop}%
\bibitem [{\citenamefont {Abi}\ \emph {et~al.}(2021)\citenamefont {Abi} \emph
  {et~al.}}]{Muong-2:2021ojo}%
  \BibitemOpen
  \bibfield  {author} {\bibinfo {author} {\bibfnamefont {B.}~\bibnamefont
  {Abi}} \emph {et~al.} (\bibinfo {collaboration} {Muon g-2 Collaboration}),\
  }\href {\doibase 10.1103/PhysRevLett.126.141801} {\bibfield  {journal}
  {\bibinfo  {journal} {Phys. Rev. Lett.}\ }\textbf {\bibinfo {volume} {126}},\
  \bibinfo {pages} {141801} (\bibinfo {year} {2021})},\ \Eprint
  {http://arxiv.org/abs/2104.03281} {arXiv:2104.03281 [hep-ex]} \BibitemShut
  {NoStop}%
\bibitem [{\citenamefont {Hou}\ and\ \citenamefont
  {Kumar}(2021)}]{Hou:2021qmf}%
  \BibitemOpen
  \bibfield  {author} {\bibinfo {author} {\bibfnamefont {W.-S.}\ \bibnamefont
  {Hou}}\ and\ \bibinfo {author} {\bibfnamefont {G.}~\bibnamefont {Kumar}},\
  }\href {\doibase 10.1140/epjc/s10052-021-09939-3} {\bibfield  {journal}
  {\bibinfo  {journal} {Eur. Phys. J. C}\ }\textbf {\bibinfo {volume} {81}},\
  \bibinfo {pages} {1132} (\bibinfo {year} {2021})},\ \Eprint
  {http://arxiv.org/abs/2107.14114} {arXiv:2107.14114 [hep-ph]} \BibitemShut
  {NoStop}%
\bibitem [{\citenamefont {Arndt}\ \emph {et~al.}(2021)\citenamefont {Arndt}
  \emph {et~al.}}]{Arndt:2021mu3etdr}%
  \BibitemOpen
  \bibfield  {author} {\bibinfo {author} {\bibfnamefont {K.}~\bibnamefont
  {Arndt}} \emph {et~al.},\ }\href {\doibase 10.1016/j.nima.2021.165679}
  {\bibfield  {journal} {\bibinfo  {journal} {Nucl. Instrum. Meth. A}\ }\textbf
  {\bibinfo {volume} {1014}},\ \bibinfo {pages} {165679} (\bibinfo {year}
  {2021})},\ \Eprint {http://arxiv.org/abs/2009.11690} {arXiv:2009.11690
  [physics.ins-det]} \BibitemShut {NoStop}%
\bibitem [{\citenamefont {Blondel}\ \emph {et~al.}(2013)\citenamefont {Blondel}
  \emph {et~al.}}]{Blondel:2013ia}%
  \BibitemOpen
  \bibfield  {author} {\bibinfo {author} {\bibfnamefont {A.}~\bibnamefont
  {Blondel}} \emph {et~al.},\ }\href@noop {} {\  (\bibinfo {year} {2013})},\
  \Eprint {http://arxiv.org/abs/1301.6113} {arXiv:1301.6113 [physics.ins-det]}
  \BibitemShut {NoStop}%
\bibitem [{\citenamefont {Kuno}\ and\ \citenamefont
  {Okada}(2001)}]{Kuno:1999jp}%
  \BibitemOpen
  \bibfield  {author} {\bibinfo {author} {\bibfnamefont {Y.}~\bibnamefont
  {Kuno}}\ and\ \bibinfo {author} {\bibfnamefont {Y.}~\bibnamefont {Okada}},\
  }\href {\doibase 10.1103/RevModPhys.73.151} {\bibfield  {journal} {\bibinfo
  {journal} {Rev. Mod. Phys.}\ }\textbf {\bibinfo {volume} {73}},\ \bibinfo
  {pages} {151} (\bibinfo {year} {2001})},\ \Eprint
  {http://arxiv.org/abs/hep-ph/9909265} {arXiv:hep-ph/9909265} \BibitemShut
  {NoStop}%
\bibitem [{\citenamefont {Camalich}\ and\ \citenamefont
  {Patel}(2022)}]{CAMALICH20221}%
  \BibitemOpen
  \bibfield  {author} {\bibinfo {author} {\bibfnamefont {J.~M.}\ \bibnamefont
  {Camalich}}\ and\ \bibinfo {author} {\bibfnamefont {M.}~\bibnamefont
  {Patel}},\ }\href {\doibase https://doi.org/10.1016/j.scib.2021.09.012}
  {\bibfield  {journal} {\bibinfo  {journal} {Science Bulletin}\ }\textbf
  {\bibinfo {volume} {67}},\ \bibinfo {pages} {1} (\bibinfo {year}
  {2022})}\BibitemShut {NoStop}%
\bibitem [{\citenamefont {Huang}\ \emph {et~al.}(2018)\citenamefont {Huang},
  \citenamefont {P\"as},\ and\ \citenamefont {Zeissner}}]{Huang:2018guts}%
  \BibitemOpen
  \bibfield  {author} {\bibinfo {author} {\bibfnamefont {W.-C.}\ \bibnamefont
  {Huang}}, \bibinfo {author} {\bibfnamefont {H.}~\bibnamefont {P\"as}}, \ and\
  \bibinfo {author} {\bibfnamefont {S.}~\bibnamefont {Zeissner}},\ }\href
  {\doibase 10.1103/PhysRevD.97.055040} {\bibfield  {journal} {\bibinfo
  {journal} {Phys. Rev. D}\ }\textbf {\bibinfo {volume} {97}},\ \bibinfo
  {pages} {055040} (\bibinfo {year} {2018})},\ \Eprint
  {http://arxiv.org/abs/1608.04354} {arXiv:1608.04354 [hep-ph]} \BibitemShut
  {NoStop}%
\bibitem [{\citenamefont {Chang}\ and\ \citenamefont
  {Feng}(2000)}]{Chang:2000sysu}%
  \BibitemOpen
  \bibfield  {author} {\bibinfo {author} {\bibfnamefont {C.-H.}\ \bibnamefont
  {Chang}}\ and\ \bibinfo {author} {\bibfnamefont {T.-F.}\ \bibnamefont
  {Feng}},\ }\href {\doibase 10.1007/s100529900194} {\bibfield  {journal}
  {\bibinfo  {journal} {Eur. Phys. J. C}\ }\textbf {\bibinfo {volume} {12}},\
  \bibinfo {pages} {137} (\bibinfo {year} {2000})},\ \Eprint
  {http://arxiv.org/abs/hep-ph/9901260} {arXiv:hep-ph/9901260} \BibitemShut
  {NoStop}%
\bibitem [{\citenamefont {Cvetic}(1999)}]{Cvetic:1999tc}%
  \BibitemOpen
  \bibfield  {author} {\bibinfo {author} {\bibfnamefont {G.}~\bibnamefont
  {Cvetic}},\ }\href {\doibase 10.1103/RevModPhys.71.513} {\bibfield  {journal}
  {\bibinfo  {journal} {Rev. Mod. Phys.}\ }\textbf {\bibinfo {volume} {71}},\
  \bibinfo {pages} {513} (\bibinfo {year} {1999})},\ \Eprint
  {http://arxiv.org/abs/hep-ph/9702381} {arXiv:hep-ph/9702381} \BibitemShut
  {NoStop}%
\bibitem [{\citenamefont {Nussinov}\ \emph {et~al.}(2001)\citenamefont
  {Nussinov}, \citenamefont {Peccei},\ and\ \citenamefont
  {Zhang}}]{Nussinov:2000nm}%
  \BibitemOpen
  \bibfield  {author} {\bibinfo {author} {\bibfnamefont {S.}~\bibnamefont
  {Nussinov}}, \bibinfo {author} {\bibfnamefont {R.~D.}\ \bibnamefont
  {Peccei}}, \ and\ \bibinfo {author} {\bibfnamefont {X.~M.}\ \bibnamefont
  {Zhang}},\ }\href {\doibase 10.1103/PhysRevD.63.016003} {\bibfield  {journal}
  {\bibinfo  {journal} {Phys. Rev. D}\ }\textbf {\bibinfo {volume} {63}},\
  \bibinfo {pages} {016003} (\bibinfo {year} {2001})},\ \Eprint
  {http://arxiv.org/abs/hep-ph/0004153} {arXiv:hep-ph/0004153} \BibitemShut
  {NoStop}%
\bibitem [{\citenamefont {Gutsche}\ \emph {et~al.}(2011)\citenamefont
  {Gutsche}, \citenamefont {Helo}, \citenamefont {Kovalenko},\ and\
  \citenamefont {Lyubovitskij}}]{Gutsche:2011bi}%
  \BibitemOpen
  \bibfield  {author} {\bibinfo {author} {\bibfnamefont {T.}~\bibnamefont
  {Gutsche}}, \bibinfo {author} {\bibfnamefont {J.~C.}\ \bibnamefont {Helo}},
  \bibinfo {author} {\bibfnamefont {S.}~\bibnamefont {Kovalenko}}, \ and\
  \bibinfo {author} {\bibfnamefont {V.~E.}\ \bibnamefont {Lyubovitskij}},\
  }\href {\doibase 10.1103/PhysRevD.83.115015} {\bibfield  {journal} {\bibinfo
  {journal} {Phys. Rev. D}\ }\textbf {\bibinfo {volume} {83}},\ \bibinfo
  {pages} {115015} (\bibinfo {year} {2011})},\ \Eprint
  {http://arxiv.org/abs/1103.1317} {arXiv:1103.1317 [hep-ph]} \BibitemShut
  {NoStop}%
\bibitem [{\citenamefont {Bordes}\ \emph {et~al.}(2001)\citenamefont {Bordes},
  \citenamefont {Chan},\ and\ \citenamefont {Tsou}}]{Bordes:2000gd}%
  \BibitemOpen
  \bibfield  {author} {\bibinfo {author} {\bibfnamefont {J.}~\bibnamefont
  {Bordes}}, \bibinfo {author} {\bibfnamefont {H.-M.}\ \bibnamefont {Chan}}, \
  and\ \bibinfo {author} {\bibfnamefont {S.~T.}\ \bibnamefont {Tsou}},\ }\href
  {\doibase 10.1103/PhysRevD.63.016006} {\bibfield  {journal} {\bibinfo
  {journal} {Phys. Rev. D}\ }\textbf {\bibinfo {volume} {63}},\ \bibinfo
  {pages} {016006} (\bibinfo {year} {2001})},\ \Eprint
  {http://arxiv.org/abs/hep-ph/0006338} {arXiv:hep-ph/0006338} \BibitemShut
  {NoStop}%
\bibitem [{\citenamefont {Sun}\ \emph {et~al.}(2012)\citenamefont {Sun} \emph
  {et~al.}}]{Sun:2012hb}%
  \BibitemOpen
  \bibfield  {author} {\bibinfo {author} {\bibfnamefont {K.-S.}\ \bibnamefont
  {Sun}} \emph {et~al.},\ }\href {\doibase 10.1142/S0217732312501726}
  {\bibfield  {journal} {\bibinfo  {journal} {Mod. Phys. Lett. A}\ }\textbf
  {\bibinfo {volume} {27}},\ \bibinfo {pages} {1250172} (\bibinfo {year}
  {2012})},\ \Eprint {http://arxiv.org/abs/1312.2072} {arXiv:1312.2072
  [hep-ph]} \BibitemShut {NoStop}%
\bibitem [{\citenamefont {Hazard}\ and\ \citenamefont
  {Petrov}(2016)}]{Hazard:2016fnc}%
  \BibitemOpen
  \bibfield  {author} {\bibinfo {author} {\bibfnamefont {D.~E.}\ \bibnamefont
  {Hazard}}\ and\ \bibinfo {author} {\bibfnamefont {A.~A.}\ \bibnamefont
  {Petrov}},\ }\href {\doibase 10.1103/PhysRevD.94.074023} {\bibfield
  {journal} {\bibinfo  {journal} {Phys. Rev. D}\ }\textbf {\bibinfo {volume}
  {94}},\ \bibinfo {pages} {074023} (\bibinfo {year} {2016})},\ \Eprint
  {http://arxiv.org/abs/1607.00815} {arXiv:1607.00815 [hep-ph]} \BibitemShut
  {NoStop}%
\bibitem [{\citenamefont {Dong}\ \emph {et~al.}(2018)\citenamefont {Dong} \emph
  {et~al.}}]{Dong:2017ipa}%
  \BibitemOpen
  \bibfield  {author} {\bibinfo {author} {\bibfnamefont {X.-X.}\ \bibnamefont
  {Dong}} \emph {et~al.},\ }\href {\doibase 10.1103/PhysRevD.97.056027}
  {\bibfield  {journal} {\bibinfo  {journal} {Phys. Rev. D}\ }\textbf {\bibinfo
  {volume} {97}},\ \bibinfo {pages} {056027} (\bibinfo {year} {2018})},\
  \Eprint {http://arxiv.org/abs/1710.07408} {arXiv:1710.07408 [hep-ph]}
  \BibitemShut {NoStop}%
\bibitem [{\citenamefont {Baldini}\ \emph {et~al.}(2016)\citenamefont {Baldini}
  \emph {et~al.}}]{meg:2016}%
  \BibitemOpen
  \bibfield  {author} {\bibinfo {author} {\bibfnamefont {A.~M.}\ \bibnamefont
  {Baldini}} \emph {et~al.} (\bibinfo {collaboration} {MEG Collaboration}),\
  }\href {\doibase 10.1140/epjc/s10052-016-4271-x} {\bibfield  {journal}
  {\bibinfo  {journal} {Eur. Phys. J. C}\ }\textbf {\bibinfo {volume} {76}},\
  \bibinfo {pages} {434} (\bibinfo {year} {2016})},\ \Eprint
  {http://arxiv.org/abs/1605.05081} {arXiv:1605.05081 [hep-ex]} \BibitemShut
  {NoStop}%
\bibitem [{\citenamefont {Aubert}\ \emph {et~al.}(2010)\citenamefont {Aubert}
  \emph {et~al.}}]{babar:2010}%
  \BibitemOpen
  \bibfield  {author} {\bibinfo {author} {\bibfnamefont {B.}~\bibnamefont
  {Aubert}} \emph {et~al.} (\bibinfo {collaboration} {BaBar Collaboration}),\
  }\href {\doibase 10.1103/PhysRevLett.104.021802} {\bibfield  {journal}
  {\bibinfo  {journal} {Phys. Rev. Lett.}\ }\textbf {\bibinfo {volume} {104}},\
  \bibinfo {pages} {021802} (\bibinfo {year} {2010})},\ \Eprint
  {http://arxiv.org/abs/0908.2381} {arXiv:0908.2381 [hep-ex]} \BibitemShut
  {NoStop}%
\bibitem [{\citenamefont {Bellgardt}\ \emph {et~al.}(1988)\citenamefont
  {Bellgardt} \emph {et~al.}}]{sind1988}%
  \BibitemOpen
  \bibfield  {author} {\bibinfo {author} {\bibfnamefont {U.}~\bibnamefont
  {Bellgardt}} \emph {et~al.} (\bibinfo {collaboration} {SINDRUM
  Collaboration}),\ }\href {\doibase 10.1016/0550-3213(88)90462-2} {\bibfield
  {journal} {\bibinfo  {journal} {Nucl. Phys. B}\ }\textbf {\bibinfo {volume}
  {299}},\ \bibinfo {pages} {1} (\bibinfo {year} {1988})}\BibitemShut {NoStop}%
\bibitem [{\citenamefont {Abouzaid}\ \emph {et~al.}(2008)\citenamefont
  {Abouzaid} \emph {et~al.}}]{KTeV:2008}%
  \BibitemOpen
  \bibfield  {author} {\bibinfo {author} {\bibfnamefont {E.}~\bibnamefont
  {Abouzaid}} \emph {et~al.} (\bibinfo {collaboration} {KTeV Collaboration}),\
  }\href {\doibase 10.1103/PhysRevLett.100.131803} {\bibfield  {journal}
  {\bibinfo  {journal} {Phys. Rev. Lett.}\ }\textbf {\bibinfo {volume} {100}},\
  \bibinfo {pages} {131803} (\bibinfo {year} {2008})},\ \Eprint
  {http://arxiv.org/abs/0711.3472} {arXiv:0711.3472 [hep-ex]} \BibitemShut
  {NoStop}%
\bibitem [{\citenamefont {Ambrose}\ \emph {et~al.}(1998)\citenamefont {Ambrose}
  \emph {et~al.}}]{BNL:1998}%
  \BibitemOpen
  \bibfield  {author} {\bibinfo {author} {\bibfnamefont {D.}~\bibnamefont
  {Ambrose}} \emph {et~al.} (\bibinfo {collaboration} {E871 Collaboration}),\
  }\href {\doibase 10.1103/PhysRevLett.81.5734} {\bibfield  {journal} {\bibinfo
   {journal} {Phys. Rev. Lett.}\ }\textbf {\bibinfo {volume} {81}},\ \bibinfo
  {pages} {5734} (\bibinfo {year} {1998})},\ \Eprint
  {http://arxiv.org/abs/hep-ex/9811038} {arXiv:hep-ex/9811038} \BibitemShut
  {NoStop}%
\bibitem [{\citenamefont {Aaij}\ \emph {et~al.}(2018)\citenamefont {Aaij} \emph
  {et~al.}}]{lhcb:2018}%
  \BibitemOpen
  \bibfield  {author} {\bibinfo {author} {\bibfnamefont {R.}~\bibnamefont
  {Aaij}} \emph {et~al.} (\bibinfo {collaboration} {LHCb Collaboration}),\
  }\href {\doibase 10.1007/JHEP03(2018)078} {\bibfield  {journal} {\bibinfo
  {journal} {JHEP}\ }\textbf {\bibinfo {volume} {03}},\ \bibinfo {pages} {078}
  (\bibinfo {year} {2018})},\ \Eprint {http://arxiv.org/abs/1710.04111}
  {arXiv:1710.04111 [hep-ex]} \BibitemShut {NoStop}%
\bibitem [{\citenamefont {Aad}\ \emph {et~al.}(2014)\citenamefont {Aad} \emph
  {et~al.}}]{ATLAS:2014vur}%
  \BibitemOpen
  \bibfield  {author} {\bibinfo {author} {\bibfnamefont {G.}~\bibnamefont
  {Aad}} \emph {et~al.} (\bibinfo {collaboration} {ATLAS Collaboration}),\
  }\href {\doibase 10.1103/PhysRevD.90.072010} {\bibfield  {journal} {\bibinfo
  {journal} {Phys. Rev. D}\ }\textbf {\bibinfo {volume} {90}},\ \bibinfo
  {pages} {072010} (\bibinfo {year} {2014})},\ \Eprint
  {http://arxiv.org/abs/1408.5774} {arXiv:1408.5774 [hep-ex]} \BibitemShut
  {NoStop}%
\bibitem [{\citenamefont {Akers}\ \emph {et~al.}(1995)\citenamefont {Akers}
  \emph {et~al.}}]{OPAL:1995grn}%
  \BibitemOpen
  \bibfield  {author} {\bibinfo {author} {\bibfnamefont {R.}~\bibnamefont
  {Akers}} \emph {et~al.} (\bibinfo {collaboration} {OPAL Collaboration}),\
  }\href {\doibase 10.1007/BF01553981} {\bibfield  {journal} {\bibinfo
  {journal} {Z. Phys. C}\ }\textbf {\bibinfo {volume} {67}},\ \bibinfo {pages}
  {555} (\bibinfo {year} {1995})}\BibitemShut {NoStop}%
\bibitem [{\citenamefont {Abreu}\ \emph {et~al.}(1997)\citenamefont {Abreu}
  \emph {et~al.}}]{DELPHI:1996iox}%
  \BibitemOpen
  \bibfield  {author} {\bibinfo {author} {\bibfnamefont {P.}~\bibnamefont
  {Abreu}} \emph {et~al.} (\bibinfo {collaboration} {DELPHI Collaboration}),\
  }\href {\doibase 10.1007/s002880050313} {\bibfield  {journal} {\bibinfo
  {journal} {Z. Phys. C}\ }\textbf {\bibinfo {volume} {73}},\ \bibinfo {pages}
  {243} (\bibinfo {year} {1997})}\BibitemShut {NoStop}%
\bibitem [{\citenamefont {Aad}\ \emph {et~al.}(2020{\natexlab{a}})\citenamefont
  {Aad} \emph {et~al.}}]{ATLAS:2019pmk}%
  \BibitemOpen
  \bibfield  {author} {\bibinfo {author} {\bibfnamefont {G.}~\bibnamefont
  {Aad}} \emph {et~al.} (\bibinfo {collaboration} {ATLAS Collaboration}),\
  }\href {\doibase 10.1016/j.physletb.2019.135069} {\bibfield  {journal}
  {\bibinfo  {journal} {Phys. Lett. B}\ }\textbf {\bibinfo {volume} {800}},\
  \bibinfo {pages} {135069} (\bibinfo {year} {2020}{\natexlab{a}})},\ \Eprint
  {http://arxiv.org/abs/1907.06131} {arXiv:1907.06131 [hep-ex]} \BibitemShut
  {NoStop}%
\bibitem [{\citenamefont {Sirunyan}\ \emph {et~al.}(2018)\citenamefont
  {Sirunyan} \emph {et~al.}}]{CMS:2017con}%
  \BibitemOpen
  \bibfield  {author} {\bibinfo {author} {\bibfnamefont {A.~M.}\ \bibnamefont
  {Sirunyan}} \emph {et~al.} (\bibinfo {collaboration} {CMS Collaboration}),\
  }\href {\doibase 10.1007/JHEP06(2018)001} {\bibfield  {journal} {\bibinfo
  {journal} {JHEP}\ }\textbf {\bibinfo {volume} {06}},\ \bibinfo {pages} {001}
  (\bibinfo {year} {2018})},\ \Eprint {http://arxiv.org/abs/1712.07173}
  {arXiv:1712.07173 [hep-ex]} \BibitemShut {NoStop}%
\bibitem [{\citenamefont {Aad}\ \emph {et~al.}(2020{\natexlab{b}})\citenamefont
  {Aad} \emph {et~al.}}]{ATLAS:2019old}%
  \BibitemOpen
  \bibfield  {author} {\bibinfo {author} {\bibfnamefont {G.}~\bibnamefont
  {Aad}} \emph {et~al.} (\bibinfo {collaboration} {ATLAS Collaboration}),\
  }\href {\doibase 10.1016/j.physletb.2019.135148} {\bibfield  {journal}
  {\bibinfo  {journal} {Phys. Lett. B}\ }\textbf {\bibinfo {volume} {801}},\
  \bibinfo {pages} {135148} (\bibinfo {year} {2020}{\natexlab{b}})},\ \Eprint
  {http://arxiv.org/abs/1909.10235} {arXiv:1909.10235 [hep-ex]} \BibitemShut
  {NoStop}%
\bibitem [{\citenamefont {Achasov}\ \emph {et~al.}(2010)\citenamefont {Achasov}
  \emph {et~al.}}]{SND:2010}%
  \BibitemOpen
  \bibfield  {author} {\bibinfo {author} {\bibfnamefont {M.~N.}\ \bibnamefont
  {Achasov}} \emph {et~al.},\ }\href {\doibase 10.1103/PhysRevD.81.057102}
  {\bibfield  {journal} {\bibinfo  {journal} {Phys. Rev. D}\ }\textbf {\bibinfo
  {volume} {81}},\ \bibinfo {pages} {057102} (\bibinfo {year} {2010})},\
  \Eprint {http://arxiv.org/abs/0911.1232} {arXiv:0911.1232 [hep-ex]}
  \BibitemShut {NoStop}%
\bibitem [{\citenamefont {Love}\ \emph {et~al.}(2008)\citenamefont {Love} \emph
  {et~al.}}]{cleo3:2008}%
  \BibitemOpen
  \bibfield  {author} {\bibinfo {author} {\bibfnamefont {W.}~\bibnamefont
  {Love}} \emph {et~al.} (\bibinfo {collaboration} {CLEO Collaboration}),\
  }\href {\doibase 10.1103/PhysRevLett.101.201601} {\bibfield  {journal}
  {\bibinfo  {journal} {Phys. Rev. Lett.}\ }\textbf {\bibinfo {volume} {101}},\
  \bibinfo {pages} {201601} (\bibinfo {year} {2008})},\ \Eprint
  {http://arxiv.org/abs/0807.2695} {arXiv:0807.2695 [hep-ex]} \BibitemShut
  {NoStop}%
\bibitem [{\citenamefont {Lees}\ \emph {et~al.}(2021)\citenamefont {Lees} \emph
  {et~al.}}]{BaBar:2021loj}%
  \BibitemOpen
  \bibfield  {author} {\bibinfo {author} {\bibfnamefont {J.~P.}\ \bibnamefont
  {Lees}} \emph {et~al.} (\bibinfo {collaboration} {BaBar Collaboration}),\
  }\href@noop {} {\  (\bibinfo {year} {2021})},\ \Eprint
  {http://arxiv.org/abs/2109.03364} {arXiv:2109.03364 [hep-ex]} \BibitemShut
  {NoStop}%
\bibitem [{\citenamefont {Patra}\ \emph {et~al.}(2022)\citenamefont {Patra}
  \emph {et~al.}}]{belle:2021}%
  \BibitemOpen
  \bibfield  {author} {\bibinfo {author} {\bibfnamefont {S.}~\bibnamefont
  {Patra}} \emph {et~al.} (\bibinfo {collaboration} {Belle Collaboration}),\
  }\href@noop {} {\  (\bibinfo {year} {2022})},\ \Eprint
  {http://arxiv.org/abs/2201.09620} {arXiv:2201.09620 [hep-ex]} \BibitemShut
  {NoStop}%
\bibitem [{\citenamefont {Ablikim}\ \emph {et~al.}(2004)\citenamefont {Ablikim}
  \emph {et~al.}}]{bes:2004}%
  \BibitemOpen
  \bibfield  {author} {\bibinfo {author} {\bibfnamefont {M.}~\bibnamefont
  {Ablikim}} \emph {et~al.} (\bibinfo {collaboration} {BES Collaboration}),\
  }\href {\doibase 10.1016/j.physletb.2004.08.005} {\bibfield  {journal}
  {\bibinfo  {journal} {Phys. Lett. B}\ }\textbf {\bibinfo {volume} {598}},\
  \bibinfo {pages} {172} (\bibinfo {year} {2004})},\ \Eprint
  {http://arxiv.org/abs/hep-ex/0406018} {arXiv:hep-ex/0406018} \BibitemShut
  {NoStop}%
\bibitem [{\citenamefont {Ablikim}\ \emph {et~al.}(2013)\citenamefont {Ablikim}
  \emph {et~al.}}]{bes3:2013}%
  \BibitemOpen
  \bibfield  {author} {\bibinfo {author} {\bibfnamefont {M.}~\bibnamefont
  {Ablikim}} \emph {et~al.} (\bibinfo {collaboration} {BESIII Collaboration}),\
  }\href {\doibase 10.1103/PhysRevD.87.112007} {\bibfield  {journal} {\bibinfo
  {journal} {Phys. Rev. D}\ }\textbf {\bibinfo {volume} {87}},\ \bibinfo
  {pages} {112007} (\bibinfo {year} {2013})},\ \Eprint
  {http://arxiv.org/abs/1304.3205} {arXiv:1304.3205 [hep-ex]} \BibitemShut
  {NoStop}%
\bibitem [{\citenamefont {Ablikim}\ \emph
  {et~al.}(2021{\natexlab{a}})\citenamefont {Ablikim} \emph
  {et~al.}}]{BESIII:2021slj}%
  \BibitemOpen
  \bibfield  {author} {\bibinfo {author} {\bibfnamefont {M.}~\bibnamefont
  {Ablikim}} \emph {et~al.} (\bibinfo {collaboration} {BESIII Collaboration}),\
  }\href {\doibase 10.1103/PhysRevD.103.112007} {\bibfield  {journal} {\bibinfo
   {journal} {Phys. Rev. D}\ }\textbf {\bibinfo {volume} {103}},\ \bibinfo
  {pages} {112007} (\bibinfo {year} {2021}{\natexlab{a}})},\ \Eprint
  {http://arxiv.org/abs/2103.11540} {arXiv:2103.11540 [hep-ex]} \BibitemShut
  {NoStop}%
\bibitem [{\citenamefont {Ablikim}\ \emph {et~al.}(2010)\citenamefont {Ablikim}
  \emph {et~al.}}]{Ablikim:2009aa}%
  \BibitemOpen
  \bibfield  {author} {\bibinfo {author} {\bibfnamefont {M.}~\bibnamefont
  {Ablikim}} \emph {et~al.} (\bibinfo {collaboration} {BESIII Collaboration}),\
  }\href {\doibase 10.1016/j.nima.2009.12.050} {\bibfield  {journal} {\bibinfo
  {journal} {Nucl. Instrum. Meth. A}\ }\textbf {\bibinfo {volume} {614}},\
  \bibinfo {pages} {345} (\bibinfo {year} {2010})},\ \Eprint
  {http://arxiv.org/abs/0911.4960} {arXiv:0911.4960 [physics.ins-det]}
  \BibitemShut {NoStop}%
\bibitem [{\citenamefont {Yu}\ \emph {et~al.}(2016)\citenamefont {Yu} \emph
  {et~al.}}]{Yu:IPAC2016-TUYA01}%
  \BibitemOpen
  \bibfield  {author} {\bibinfo {author} {\bibfnamefont {C.}~\bibnamefont {Yu}}
  \emph {et~al.},\ }in\ \href {\doibase 10.18429/JACoW-IPAC2016-TUYA01} {\emph
  {\bibinfo {booktitle} {{7th International Particle Accelerator
  Conference}}}}\ (\bibinfo {year} {2016})\ p.\ \bibinfo {pages}
  {TUYA01}\BibitemShut {NoStop}%
\bibitem [{\citenamefont {Ablikim}\ \emph {et~al.}(2020)\citenamefont {Ablikim}
  \emph {et~al.}}]{BESIII:2020nme}%
  \BibitemOpen
  \bibfield  {author} {\bibinfo {author} {\bibfnamefont {M.}~\bibnamefont
  {Ablikim}} \emph {et~al.} (\bibinfo {collaboration} {BESIII Collaboration}),\
  }\href {\doibase 10.1088/1674-1137/44/4/040001} {\bibfield  {journal}
  {\bibinfo  {journal} {Chin. Phys. C}\ }\textbf {\bibinfo {volume} {44}},\
  \bibinfo {pages} {040001} (\bibinfo {year} {2020})},\ \Eprint
  {http://arxiv.org/abs/1912.05983} {arXiv:1912.05983 [hep-ex]} \BibitemShut
  {NoStop}%
\bibitem [{\citenamefont {Li}\ \emph {et~al.}(2017)\citenamefont {Li} \emph
  {et~al.}}]{etof1}%
  \BibitemOpen
  \bibfield  {author} {\bibinfo {author} {\bibfnamefont {X.}~\bibnamefont {Li}}
  \emph {et~al.},\ }\href {\doibase 10.1007/s41605-017-0014-2} {\bibfield
  {journal} {\bibinfo  {journal} {Radiation Detection Technology and Methods}\
  }\textbf {\bibinfo {volume} {1}} (\bibinfo {year} {2017}),\
  10.1007/s41605-017-0014-2}\BibitemShut {NoStop}%
\bibitem [{\citenamefont {Guo}\ \emph {et~al.}(2017)\citenamefont {Guo} \emph
  {et~al.}}]{etof2}%
  \BibitemOpen
  \bibfield  {author} {\bibinfo {author} {\bibfnamefont {Y.-X.}\ \bibnamefont
  {Guo}} \emph {et~al.},\ }\href {\doibase 10.1007/s41605-017-0012-4}
  {\bibfield  {journal} {\bibinfo  {journal} {Radiation Detection Technology
  and Methods}\ }\textbf {\bibinfo {volume} {1}} (\bibinfo {year} {2017}),\
  10.1007/s41605-017-0012-4}\BibitemShut {NoStop}%
\bibitem [{\citenamefont {Ablikim}\ \emph {et~al.}(2017)\citenamefont {Ablikim}
  \emph {et~al.}}]{bes3:totJpsiNumber}%
  \BibitemOpen
  \bibfield  {author} {\bibinfo {author} {\bibfnamefont {M.}~\bibnamefont
  {Ablikim}} \emph {et~al.} (\bibinfo {collaboration} {BESIII Collaboration}),\
  }\href {\doibase 10.1088/1674-1137/41/1/013001} {\bibfield  {journal}
  {\bibinfo  {journal} {Chin. Phys. C}\ }\textbf {\bibinfo {volume} {41}},\
  \bibinfo {pages} {013001} (\bibinfo {year} {2017})},\ \Eprint
  {http://arxiv.org/abs/1607.00738} {arXiv:1607.00738 [hep-ex]} \BibitemShut
  {NoStop}%
\bibitem [{\citenamefont {Ablikim}\ \emph
  {et~al.}(2021{\natexlab{b}})\citenamefont {Ablikim} \emph
  {et~al.}}]{bes3:totJpsiNumber2}%
  \BibitemOpen
  \bibfield  {author} {\bibinfo {author} {\bibfnamefont {M.}~\bibnamefont
  {Ablikim}} \emph {et~al.} (\bibinfo {collaboration} {BESIII Collaboration}),\
  }\href@noop {} {\  (\bibinfo {year} {2021}{\natexlab{b}})},\ \Eprint
  {http://arxiv.org/abs/2111.07571} {arXiv:2111.07571 [hep-ex]} \BibitemShut
  {NoStop}%
\bibitem [{\citenamefont {Agostinelli}\ \emph {et~al.}(2003)\citenamefont
  {Agostinelli} \emph {et~al.}}]{geant4}%
  \BibitemOpen
  \bibfield  {author} {\bibinfo {author} {\bibfnamefont {S.}~\bibnamefont
  {Agostinelli}} \emph {et~al.} (\bibinfo {collaboration} {GEANT4
  Collaboration}),\ }\href {\doibase 10.1016/S0168-9002(03)01368-8} {\bibfield
  {journal} {\bibinfo  {journal} {Nucl. Instrum. Meth. A}\ }\textbf {\bibinfo
  {volume} {506}},\ \bibinfo {pages} {250} (\bibinfo {year}
  {2003})}\BibitemShut {NoStop}%
\bibitem [{\citenamefont {Deng}\ and\ \citenamefont {the
  others}(2006)\citenamefont {Deng} \emph {et~al.}}]{bes:boost}%
  \BibitemOpen
  \bibfield  {author} {\bibinfo {author} {\bibfnamefont {Z.-Y.}\ \bibnamefont
  {Deng}} \emph {et~al.},\ }\href@noop {} {\bibfield  {journal} {\bibinfo
  {journal} {Chin. Phys. C}\ }\textbf {\bibinfo {volume} {30}},\ \bibinfo
  {pages} {371} (\bibinfo {year} {2006})}\BibitemShut {NoStop}%
\bibitem [{\citenamefont {Liang}\ \emph {et~al.}(2009)\citenamefont {Liang}
  \emph {et~al.}}]{geo1}%
  \BibitemOpen
  \bibfield  {author} {\bibinfo {author} {\bibfnamefont {Y.-T.}\ \bibnamefont
  {Liang}} \emph {et~al.},\ }\href {\doibase 10.1016/j.nima.2009.02.036}
  {\bibfield  {journal} {\bibinfo  {journal} {Nucl. Instrum. Meth. A}\ }\textbf
  {\bibinfo {volume} {603}},\ \bibinfo {pages} {325} (\bibinfo {year}
  {2009})}\BibitemShut {NoStop}%
\bibitem [{\citenamefont {You}\ \emph {et~al.}(2008)\citenamefont {You},
  \citenamefont {Liang},\ and\ \citenamefont {Mao}}]{geo2}%
  \BibitemOpen
  \bibfield  {author} {\bibinfo {author} {\bibfnamefont {Z.-Y.}\ \bibnamefont
  {You}}, \bibinfo {author} {\bibfnamefont {Y.-T.}\ \bibnamefont {Liang}}, \
  and\ \bibinfo {author} {\bibfnamefont {Y.-J.}\ \bibnamefont {Mao}},\ }\href
  {\doibase 10.1088/1674-1137/32/7/012} {\bibfield  {journal} {\bibinfo
  {journal} {Chin. Phys. C}\ }\textbf {\bibinfo {volume} {32}},\ \bibinfo
  {pages} {572} (\bibinfo {year} {2008})}\BibitemShut {NoStop}%
\bibitem [{\citenamefont {Huang}\ \emph {et~al.}(2022)\citenamefont {Huang},
  \citenamefont {Li}, \citenamefont {Qian}, \citenamefont {Zhu}, \citenamefont
  {Li}, \citenamefont {Zhang}, \citenamefont {Sun},\ and\ \citenamefont
  {You}}]{Huang:2022wuo}%
  \BibitemOpen
  \bibfield  {author} {\bibinfo {author} {\bibfnamefont {K.-X.}\ \bibnamefont
  {Huang}}, \bibinfo {author} {\bibfnamefont {Z.-J.}\ \bibnamefont {Li}},
  \bibinfo {author} {\bibfnamefont {Z.}~\bibnamefont {Qian}}, \bibinfo {author}
  {\bibfnamefont {J.}~\bibnamefont {Zhu}}, \bibinfo {author} {\bibfnamefont
  {H.-Y.}\ \bibnamefont {Li}}, \bibinfo {author} {\bibfnamefont {Y.-M.}\
  \bibnamefont {Zhang}}, \bibinfo {author} {\bibfnamefont {S.-S.}\ \bibnamefont
  {Sun}}, \ and\ \bibinfo {author} {\bibfnamefont {Z.-Y.}\ \bibnamefont
  {You}},\ }\href {\doibase 10.1007/s41365-022-01133-8} {\bibfield  {journal}
  {\bibinfo  {journal} {Nucl. Sci. Tech.}\ }\textbf {\bibinfo {volume} {33}},\
  \bibinfo {pages} {142} (\bibinfo {year} {2022})},\ \Eprint
  {http://arxiv.org/abs/2206.10117.} {arXiv:2206.10117. [physics.ins-det]}
  \BibitemShut {NoStop}%
\bibitem [{\citenamefont {Ablikim}(2013)}]{Ablikim:2013ntc}%
  \BibitemOpen
  \bibfield  {author} {\bibinfo {author} {\bibfnamefont {M.}~\bibnamefont
  {Ablikim}} (\bibinfo {collaboration} {BESIII Collaboration}),\ }\href
  {\doibase 10.1088/1674-1137/37/12/123001} {\bibfield  {journal} {\bibinfo
  {journal} {Chin. Phys. C}\ }\textbf {\bibinfo {volume} {37}},\ \bibinfo
  {pages} {123001} (\bibinfo {year} {2013})},\ \Eprint
  {http://arxiv.org/abs/1307.2022} {arXiv:1307.2022 [hep-ex]} \BibitemShut
  {NoStop}%
\bibitem [{\citenamefont {Jadach}\ \emph {et~al.}(2001)\citenamefont {Jadach},
  \citenamefont {Ward},\ and\ \citenamefont {Was}}]{ref:kkmc1}%
  \BibitemOpen
  \bibfield  {author} {\bibinfo {author} {\bibfnamefont {S.}~\bibnamefont
  {Jadach}}, \bibinfo {author} {\bibfnamefont {B.~F.~L.}\ \bibnamefont {Ward}},
  \ and\ \bibinfo {author} {\bibfnamefont {Z.}~\bibnamefont {Was}},\ }\href
  {\doibase 10.1103/PhysRevD.63.113009} {\bibfield  {journal} {\bibinfo
  {journal} {Phys. Rev. D}\ }\textbf {\bibinfo {volume} {63}},\ \bibinfo
  {pages} {113009} (\bibinfo {year} {2001})},\ \Eprint
  {http://arxiv.org/abs/hep-ph/0006359} {arXiv:hep-ph/0006359} \BibitemShut
  {NoStop}%
\bibitem [{\citenamefont {Jadach}\ \emph {et~al.}(2000)\citenamefont {Jadach},
  \citenamefont {Ward},\ and\ \citenamefont {Was}}]{ref:kkmc2}%
  \BibitemOpen
  \bibfield  {author} {\bibinfo {author} {\bibfnamefont {S.}~\bibnamefont
  {Jadach}}, \bibinfo {author} {\bibfnamefont {B.~F.~L.}\ \bibnamefont {Ward}},
  \ and\ \bibinfo {author} {\bibfnamefont {Z.}~\bibnamefont {Was}},\ }\href
  {\doibase 10.1016/S0010-4655(00)00048-5} {\bibfield  {journal} {\bibinfo
  {journal} {Comput. Phys. Commun.}\ }\textbf {\bibinfo {volume} {130}},\
  \bibinfo {pages} {260} (\bibinfo {year} {2000})},\ \Eprint
  {http://arxiv.org/abs/hep-ph/9912214} {arXiv:hep-ph/9912214} \BibitemShut
  {NoStop}%
\bibitem [{\citenamefont {Lange}(2001)}]{ref:evtgen1}%
  \BibitemOpen
  \bibfield  {author} {\bibinfo {author} {\bibfnamefont {D.~J.}\ \bibnamefont
  {Lange}},\ }\href {\doibase 10.1016/S0168-9002(01)00089-4} {\bibfield
  {journal} {\bibinfo  {journal} {Nucl. Instrum. Meth. A}\ }\textbf {\bibinfo
  {volume} {462}},\ \bibinfo {pages} {152} (\bibinfo {year}
  {2001})}\BibitemShut {NoStop}%
\bibitem [{\citenamefont {Chen}\ \emph {et~al.}(2000)\citenamefont {Chen},
  \citenamefont {Huang}, \citenamefont {Qi}, \citenamefont {Zhang},\ and\
  \citenamefont {Zhu}}]{Chen:2000tv}%
  \BibitemOpen
  \bibfield  {author} {\bibinfo {author} {\bibfnamefont {J.~C.}\ \bibnamefont
  {Chen}}, \bibinfo {author} {\bibfnamefont {G.~S.}\ \bibnamefont {Huang}},
  \bibinfo {author} {\bibfnamefont {X.~R.}\ \bibnamefont {Qi}}, \bibinfo
  {author} {\bibfnamefont {D.~H.}\ \bibnamefont {Zhang}}, \ and\ \bibinfo
  {author} {\bibfnamefont {Y.~S.}\ \bibnamefont {Zhu}},\ }\href {\doibase
  10.1103/PhysRevD.62.034003} {\bibfield  {journal} {\bibinfo  {journal} {Phys.
  Rev. D}\ }\textbf {\bibinfo {volume} {62}},\ \bibinfo {pages} {034003}
  (\bibinfo {year} {2000})}\BibitemShut {NoStop}%
\bibitem [{\citenamefont {Yang}\ \emph {et~al.}(2014)\citenamefont {Yang},
  \citenamefont {Ping},\ and\ \citenamefont {Chen}}]{ref:lundcharm2}%
  \BibitemOpen
  \bibfield  {author} {\bibinfo {author} {\bibfnamefont {R.-L.}\ \bibnamefont
  {Yang}}, \bibinfo {author} {\bibfnamefont {R.-G.}\ \bibnamefont {Ping}}, \
  and\ \bibinfo {author} {\bibfnamefont {H.}~\bibnamefont {Chen}},\ }\href
  {\doibase 10.1088/0256-307X/31/6/061301} {\bibfield  {journal} {\bibinfo
  {journal} {Chin. Phys. Lett.}\ }\textbf {\bibinfo {volume} {31}},\ \bibinfo
  {pages} {061301} (\bibinfo {year} {2014})}\BibitemShut {NoStop}%
\bibitem [{\citenamefont {Richter-Was}(1993)}]{photos}%
  \BibitemOpen
  \bibfield  {author} {\bibinfo {author} {\bibfnamefont {E.}~\bibnamefont
  {Richter-Was}},\ }\href {\doibase 10.1016/0370-2693(93)90062-M} {\bibfield
  {journal} {\bibinfo  {journal} {Phys. Lett. B}\ }\textbf {\bibinfo {volume}
  {303}},\ \bibinfo {pages} {163} (\bibinfo {year} {1993})}\BibitemShut
  {NoStop}%
\bibitem [{\citenamefont {Li}(2006)}]{bes3:boss705}%
  \BibitemOpen
  \bibfield  {author} {\bibinfo {author} {\bibfnamefont {W.-D.}\ \bibnamefont
  {Li}},\ }\href {https://indico.cern.ch/event/408139/contributions/979815/}
  {\bibfield  {journal} {\bibinfo  {journal} {Proceeding of CHEP06, Mumbai,
  India}\ } (\bibinfo {year} {2006})}\BibitemShut {NoStop}%
\bibitem [{\citenamefont {Feldman}\ and\ \citenamefont
  {Cousins}(1998)}]{feldman:1998ul}%
  \BibitemOpen
  \bibfield  {author} {\bibinfo {author} {\bibfnamefont {G.~J.}\ \bibnamefont
  {Feldman}}\ and\ \bibinfo {author} {\bibfnamefont {R.~D.}\ \bibnamefont
  {Cousins}},\ }\href {\doibase 10.1103/PhysRevD.57.3873} {\bibfield  {journal}
  {\bibinfo  {journal} {Phys. Rev. D}\ }\textbf {\bibinfo {volume} {57}},\
  \bibinfo {pages} {3873} (\bibinfo {year} {1998})},\ \Eprint
  {http://arxiv.org/abs/physics/9711021} {arXiv:physics/9711021} \BibitemShut
  {NoStop}%
\bibitem [{\citenamefont {Rolke}\ \emph {et~al.}(2005)\citenamefont {Rolke},
  \citenamefont {Lopez},\ and\ \citenamefont {Conrad}}]{rolke:2005like}%
  \BibitemOpen
  \bibfield  {author} {\bibinfo {author} {\bibfnamefont {W.~A.}\ \bibnamefont
  {Rolke}}, \bibinfo {author} {\bibfnamefont {A.~M.}\ \bibnamefont {Lopez}}, \
  and\ \bibinfo {author} {\bibfnamefont {J.}~\bibnamefont {Conrad}},\ }\href
  {\doibase 10.1016/j.nima.2005.05.068} {\bibfield  {journal} {\bibinfo
  {journal} {Nucl. Instrum. Meth. A}\ }\textbf {\bibinfo {volume} {551}},\
  \bibinfo {pages} {493} (\bibinfo {year} {2005})},\ \Eprint
  {http://arxiv.org/abs/physics/0403059} {arXiv:physics/0403059} \BibitemShut
  {NoStop}%
\end{thebibliography}%

\vspace{0.8cm}
\noindent
\textbf{The BESIII experiment}
\vspace{0.2cm}

BESIII is an experiment located at the Beijing Electron Positron Collider (BEPCII) for the studies of hadron physics and $\tau$-charm physics. The BESIII detector consists of a helium-based multi-layer drift chamber, a plastic scintillator time-of-flight system, a CsI(Tl) electromagnetic calorimeter, and a muon identification system. The BESIII Collaboration has more than 500 members from 75 institutions in 15 countries.

\author{Author list}
\begin{small}
\vspace{0.4cm}
\noindent
\textbf{BESIII Collaboration}
\vspace{0.1cm}
\begin{center}
M.~Ablikim$^{1}$, M.~N.~Achasov$^{11,b}$, P.~Adlarson$^{70}$, M.~Albrecht$^{4}$, R.~Aliberti$^{31}$, A.~Amoroso$^{69A,69C}$, M.~R.~An$^{35}$, Q.~An$^{66,53}$, X.~H.~Bai$^{61}$, Y.~Bai$^{52}$
% ...
\\
\vspace{0.2cm} {\it
$^{1}$ Institute of High Energy Physics, Beijing 100049, People's Republic of China\\
% ...
$^{4}$ Bochum Ruhr-University, D-44780 Bochum, Germany\\
% ...
$^{11}$ G.I. Budker Institute of Nuclear Physics SB RAS (BINP), Novosibirsk 630090, Russia\\
% ...
$^{31}$ Johannes Gutenberg University of Mainz, Johann-Joachim-Becher-Weg 45, D-55099 Mainz, Germany\\
% ...
$^{35}$ Liaoning Normal University, Dalian 116029, People's Republic of China\\
% ...
$^{52}$ Southeast University, Nanjing 211100, People's Republic of China\\
$^{53}$ State Key Laboratory of Particle Detection and Electronics, Beijing 100049, Hefei 230026, People's Republic of China\\
% ...
$^{61}$ University of Jinan, Jinan 250022, People's Republic of China\\
% ...
$^{66}$ University of Science and Technology of China, Hefei 230026, People's Republic of China\\
% ...
$^{69}$ University of Turin and INFN, (A)University of Turin, I-10125, Turin, Italy; (B)University of Eastern Piedmont, I-15121, Alessandria, Italy; (C)INFN, I-10125, Turin, Italy\\
$^{70}$ Uppsala University, Box 516, SE-75120 Uppsala, Sweden\\
% ...
$^{b}$ Also at the Novosibirsk State University, Novosibirsk, 630090, Russia\\
% ...
}\end{center}

\vspace{0.4cm}
\end{small}

\end{document}